\documentclass[twocolumn,aps,prb,10pt,nofootinbib,superscriptaddress]{revtex4-2}

\usepackage{graphicx}
\usepackage{amsmath,amssymb}
\usepackage{dsfont}
\usepackage[utf8]{inputenc}
\usepackage[dvipsnames]{xcolor}
\usepackage{tabularx}
\usepackage{comment}
\usepackage[normalem]{ulem}

\usepackage[colorlinks=true,urlcolor=blue,citecolor=blue,linkcolor=blue,breaklinks=true]{hyperref}

\DeclareMathOperator{\Tr}{Tr}

\DeclareMathOperator{\e}{e}

\newcommand{\diff}[1]{\mathrm d #1\,}

\newcolumntype{L}{>{\raggedright\arraybackslash}X}

\begin{document}

\title{Theory of critical behavior at non-polar structural phase transitions from functional renormalization}

\author{Max Fornoville}

\affiliation{Max Planck Institute for Solid State Research, 70569 Stuttgart, Germany}

\affiliation{School of Natural Sciences, Technische Universit\"at M\"unchen, 85748 Garching, Germany}

\author{Kilian Fraboulet}

\affiliation{Max Planck Institute for Solid State Research, 70569 Stuttgart, Germany}

\author{Michael M. Scherer}

\affiliation{Theoretische Physik III, Ruhr-Universität Bochum, 44801 Bochum, Germany}

\author{Laura Classen}

\affiliation{Max Planck Institute for Solid State Research, 70569 Stuttgart, Germany}

\affiliation{School of Natural Sciences, Technische Universit\"at M\"unchen, 85748 Garching, Germany}

\affiliation{Munich Center for Quantum Science and Technology (MCQST), Schellingstr. 4, 80799 M\"unchen, Germany}

\begin{abstract}
$O(N)$ models provide a versatile framework for describing continuous phase transitions in a wide variety of physical systems. We investigate an extension of these models arising from order parameters that transform under interlocked internal and spatial symmetries which describe transitions towards non-polar structural order. Such order occurs, for instance, in dipolar antiferromagnets, spin-orbit-coupled magnets, non-polar ferroics, and certain nematic systems. The resulting effective theory is an $O(d)$ model with an additional symmetry-allowed kinetic term $h(\partial_i\phi_i)^2$, which modifies the relative stiffness of longitudinal and transverse fluctuations. We study both thermal and quantum phase transitions within this $O(d)$ model using the functional renormalization group (FRG). In contrast to conventional $O(N)$ models, the interlocking of internal and spatial symmetries distinguishes spatial and imaginary-time derivatives, such that the quantum model is not simply obtained from the thermal one by a dimensional extension $d\rightarrow d+1$. We derive and analyze the corresponding FRG flow equations, paying particular attention to the tensor structure of the two-point function, which is essential for consistently extracting the flow of the kinetic terms. Our analysis shows that the additional kinetic coupling is irrelevant at the $O(d)$ Wilson-Fisher fixed point in both thermal and quantum cases, where quantum-to-classical correspondence thus re-emerges. However, its small but sizable correction-to-scaling exponent gives rise to substantial corrections to scaling in the vicinity of the critical point.
\end{abstract}

\date{\today}

\maketitle

\section{Introduction}

The fascinating phenomenon of universality arises close to second-order phase transitions where the emergent behavior of microscopically very different many-body systems exhibits the same macroscopic thermodynamic properties characterized by critical exponents of scaling laws~\cite{Kadanoff1966,Wilson1971,Pelissetto2002,Herbut2007,Sachdev2011}. In the universal regime, a system can be described by effective field theories for the emergent degrees of freedom based on its symmetries rather than by a theory for its microscopic constituents~\cite{Herz1976,Goldenfeld1992,Millis1993,ZinnJustin2002,Vojta2003}. A prominent example for such an effective field theory is scalar $\phi^4$ theory with $O(N)$ symmetry, where $\phi$ describes the bosonic order-parameter field. The $O(N)$ model has played an important role for understanding second-order phase transitions in statistical and quantum systems~\cite{Pelissetto2002,ZinnJustin2002}. Its broad applicability stems from the fact that $O(N)$ symmetries and their spontaneous breaking occur in a wide variety of physical systems. Another reason is its relative simplicity so that $O(N)$ models are often used as benchmark theories for non-perturbative approaches~\cite{Coleman1974,Cant1979,Feinberg1995,ZinnJustin2002,Berges2002,Aarts2002,Moshe2003,Cooper2005,Baacke05,Delamotte2012,Keitel2012,Gergely2013,Dupuis2021,Koenigstein2022,Fraboulet2023,Yokota2024}.

However, the wide applicability of $O(N)$ models often implicitly assumes previously established results on the irrelevance of perturbations on the microscopic level such as, for example, rotational anisotropies due to the lattice of a solid-state or magnetic system~\cite{Cardy1996,Pelissetto2002}. This has to be checked for each given perturbation separately. Here, we consider systems with non-polar structural order, as opposed to polar order which refers to ordered states that preserve time reversal symmetry, while breaking spatial inversion symmetry in the usual crystallographic classification~\cite{Saxena2011}. The non-polar phase transitions that we aim to describe also break an interlocked internal and spatial symmetry. This means that $O(N)$ rotations of the order parameter and spatial rotations cannot be performed independently. Only their combination constitutes a symmetry, which immediately implies $N=d$. In addition to that, the non-polar nature of the order parameter imposes an additional internal $\mathds Z_2$ symmetry, which is broken at the phase transition. Physically, this symmetry breaking is either associated with a spontaneously broken time reversal symmetry, examples including antiferromagnetic or short-range ferromagnetic dipolar magnets, spin-orbit coupled magnets and ferrotoroidic materials~\cite{Aharony1973,Aharony1973a,Aharony1973b,Bak1980,Frey1994,Gnewuch2019,Picozzi2024}, or a preservation of spatial inversion symmetry while simultaneously breaking rotational symmetry, which is realized in ferroelastic and some ferroaxial systems, as well as nematics breaking three- or sixfold rotational symmetries~\cite{Salje1990,Salje2005,Hlinka2016,Fernandes2019,Binghai2024}.  Such phases of matter play an increasingly important role in quantum materials, where lattice, charge, orbital, and spin degrees of freedom are strongly coupled.

In systems with non-polar structural order the symmetry is lowered from independent internal and spatial $O(d)\times O(d)$ to combined $O(d) \times \mathds Z_2$ rotations. This allows an additional term in the kinetic energy of the effective theory proportional to $(\partial_i \phi_i)^2$ altering the stiffness or velocities of longitudinal and transverse modes. The additional term possesses the same canonical scaling as the conventional $(\partial_i \phi_j)^2$ kinetic term so that it is important to determine if it is relevant or irrelevant in the renormalization group (RG) sense for the phase transition.

Interestingly, if internal and spatial transformations are coupled, it also becomes necessary to distinguish thermal and quantum phase transitions beyond a classical-to-quantum mapping via dimensional extension $d\rightarrow d+1$ already on the level of the bosonic order-parameter theory. The reason is that the locking to spatial symmetries automatically breaks any potential equivalence between space and (imaginary) time coordinates. Thus, Lorentz invariance of the conventional $O(N)$ model is absent in the bare action. However, if the corresponding perturbation is irrelevant, Lorentz invariance can emerge close to the phase transition and quantum-to-classical equivalence can effectively reappear. 

While the relevance of perturbations due to interlocked internal and spatial symmetries was studied for the classical case~\cite{Aharony1973a,Aharony1973,Adzhemyan2023,Gimenez2024,Nakayama2024,Kalagov2026,Matsumoto2026}, it remains an open question in the quantum case. Originally, the classical case was motivated by dipolar antiferromagnets and short-range dipolar ferromagnets~\cite{Aharony1973a}, and the perturbation to the kinetic energy was shown to be irrelevant up to four-loop corrections in a RG approach expanding around the upper critical dimension of four~\cite{Adzhemyan2023}. This behavior should be contrasted with ferromagnets with long-ranged dipolar interactions $\propto p_i p_j/p^2$, which have a strong effect because they are relevant, so that different critical behavior arises~\cite{Aharony1973,Gimenez2024,Nakayama2024,Kalagov2026,Matsumoto2026}. Furthermore, in the short-range case, the exponent that characterizes corrections to scaling $\omega$ and thus determines the size of the scaling regime was estimated to be close to zero~\cite{Aharony1973a,Adzhemyan2023}. This implies that scaling corrections are strong and should be taken into account in any numerical treatment of such a phase transition, even though the perturbation is irrelevant. Therefore, non-perturbative calculations of $\omega$ from different methods, as well as studies of the quantum phase transition to structural order are desirable, one of which we provide here using the functional RG (FRG)~\cite{Wetterich1993,Ellwanger1994,Morris1994,Berges2002,Kopietz2010,Delamotte2012,Metzner2012,Dupuis2021}.

We consider the classical and quantum effective $O(d) \times \mathds Z_2$ field theory for a $d$-component order parameter describing a structural phase transition in $d$ spatial dimensions which includes the additional symmetry-allowed kinetic term $\propto h (\partial_i\phi_i)^2$. To study the associated critical behavior, we derive RG equations and perform a fixed-point analysis within an extended local potential approximation (LPA)~\cite{Morris1994a,Adams1995,Berges1996,Berges2002,Blaizot2006} of the FRG framework. We determine explicit expressions for the critical exponents including anomalous dimension, dynamical critical exponent, correlation-length exponent, and correction-to-scaling exponent. We find that the coupling $h$ is irrelevant at the $O(d)$ Wilson-Fisher fixed point, but show that scaling corrections are large, providing an independent confirmation of the classical result and extending it to the quantum case. On a methodological level, we show that it is crucial to account for the correct tensor structures when using projections in the derivation of flow equations for the kinetic terms, which can only be seen from higher-order derivative expansions. 

The rest of the paper is organized as follows. In Sec.~\ref{sec:Models}, we first introduce the symmetries of the $O(d) \times \mathds Z_2$ model considered in this work, explicitly defining the interlocking of its internal and spatial symmetries. From this, the classical and quantum versions of our $O(d) \times \mathds Z_2$ model are defined. In Sec.~\ref{sec:RG}, the main ingredients of our RG analysis are presented, including the definitions of the $\beta$ functions and the critical exponents, as well as the approximation scheme employed in our FRG setup. Sec.~\ref{sec:FRG} discusses our choice of regulators and presents our FRG flow equations for both the classical and quantum $O(d) \times \mathds Z_2$ models. These flow equations are then used in Sec.~\ref{sec:CriticalBehavior} to analyze the critical behavior associated with the Wilson-Fisher fixed point of both models. Finally, our concluding remarks are given in Sec.~\ref{sec:Conclusions}.

\section{Models}
\label{sec:Models}
\subsection{Symmetries}

It is instructive to first consider a classical isotropic ferromagnet in $d=3$ spatial dimensions and then build a model of interlocked internal and spatial symmetries from there. The isotropic ferromagnet is commonly described by the $O(3)$ Landau-Ginzburg-Wilson effective field theory
\begin{equation}\label{eq:standardO3}
    S = \int_\mathbf{x} 
    \left[
    \frac 12 (\partial_i \phi_j)^2 + \frac 12 m^2 \phi_i^2 + \frac \lambda 4 (\phi_i^2)^2
    \right],
\end{equation}
where $i = 1,2,3$. Here, $\phi_i$ is a three-component order parameter, which in case of a ferromagnet corresponds to the macroscopic magnetization of the system.

The above theory is invariant under both internal and spatial $O(3)$ rotations separately. This can be understood from its microscopic predecessor, the Heisenberg model, which assumes that spins on different lattice sites interact via exchange interactions only. Within this approximation, the spin degrees of freedom fully decouple from spatial rotations, thus leading to independent internal and spatial symmetries. The order parameter, which essentially arises from coarse-graining the microscopic spins, inherits this symmetry. From a group-theoretical perspective, $\phi$ transforms under the vector representation of internal $O(3)$, while being a scalar multiplet of spatial $O(3)$, giving rise to the full symmetry group $O(3) \times O(3)$.

However, in any real material there are several sources that couple internal and spatial degrees of freedom such as spin-orbit coupling or the direct dipolar interaction between two magnetic moments. In the latter case, the magnetic order parameter naturally couples to the magnetic field of the surrounding spins $E_B = -\phi_i B_i$, which mediates an effective long-range dipolar self-interaction
\begin{equation}
    V_\mathrm{dipolar} = -v \int_{\mathbf x, \mathbf y} \partial_i \partial_j \frac{1}{\vert \mathbf x - \mathbf y \vert} \phi_i(\mathbf x) \phi_j(\mathbf y).
\end{equation}
Due to this, the order parameter also transforms as a vector under spatial rotations, which now affect both internal and spatial coordinates simultaneously, effectively locking the two symmetries into a single $O(3)$ symmetry. In addition, due to time reversal symmetry, we retain a residual $\mathds Z_2$ spin-flip symmetry that leaves the spatial coordinates unaffected and only changes the sign of the order parameter. As we explain in the next section, Eq.~\eqref{eq:standardO3} needs to be adapted in such a case of interlocked $O(3)$ symmetry because additional kinetic terms are allowed to appear in the effective action.

To generalize this to arbitrary spatial dimensions $d$, we observe that, due to the interlocking of the two symmetries, the number of field components has to be fixed to $d$. Accordingly, the full symmetry of the system in arbitrary dimension becomes $O(d) \times \mathds Z_2$, with
\begin{align}
    O(d) &\mathrm : \quad x_i \mapsto O_{ij} x_j, \quad \phi_i \mapsto O_{ij} \phi_j,\\
    \mathds Z_2 &\mathrm : \quad \phi_i \mapsto -\phi_i.
\end{align}
Condensation of the order parameter then signals a symmetry reduction
\begin{equation}
    O(d) \times \mathds Z_2 \rightarrow O(d-1),
\end{equation}
signaling a spontaneous breakdown of the interlocked rotational and internal symmetry of the model.

This set of symmetries presents the outset for the remainder of this paper. In the following, we derive effective models for both thermal and quantum phase transitions of systems that spontaneously break interlocked internal and spatial symmetries. These include both antiferromagnets and ferromagnets with short-range dipolar interactions~\cite{Aharony1973a}. Here, the long-range part of the interaction is absent but retains the coupling to the magnetic field, and thus the symmetries remain intact. We want to emphasize that the following discussion assumes only the presence of $O(d) \times \mathds Z_2$ symmetry and is independent of any of the preceding microscopic considerations. As we have already mentioned in the introduction, the internal $\mathds Z_2$ symmetry appears naturally in systems describing any kind of non-polar order, assuming that the system has time-reversal and inversion symmetry. As such, our conclusions apply equally to systems without an inherent dipolar interaction, assuming that their order parameters are non-polar. This includes systems like spin-orbit-coupled magnets, non-polar ferroics and nematic systems, which spontaneously break rotational symmetry from short-range interactions~\cite{Aharony1973,Aharony1973a,Aharony1973b,Bak1980,Frey1994,Gnewuch2019,Picozzi2024,Salje1990,Salje2005,Hlinka2016,Fernandes2019,Binghai2024}.

\subsection{Classical phase transition}

In the thermal case $T > 0$ we restrict ourselves to classical fluctuations in space $\phi(\mathbf x)$ (as opposed to quantum fluctuations in space and imaginary time $\phi(\mathbf x,\tau)$). By symmetry, two independent gradient terms are allowed, which distinguishes the theory from the usual O(d) model with independent internal and spatial symmetries, cf.~Eq.~\eqref{eq:standardO3}. The two gradient terms can be deduced from the fact that the momentum-space kernel $K_{ij}(\mathbf p)$ is a rank-two symmetric tensor under $O(d)$ and thus naturally decomposes into a scalar trace $\delta_{ij}$ and a traceless symmetric tensor $s_{ij} = \tfrac{p_i p_j}{p^2} - \tfrac 1d \delta_{ij}$, giving
\begin{equation}
   \mathcal L_0= \phi_i(-\mathbf p) K_{ij}(\mathbf p) \phi_j(\mathbf p), \quad K_{ij}(\mathbf p) = \mathbf p^2(\alpha \delta_{ij} + \beta s_{ij}),
\end{equation}
with spatial indices $i,j = 1,\dots,d$. Positive definiteness of the action imposes stability bounds on the couplings $\alpha$ and $\beta$. These become apparent after a slight rewriting in terms of longitudinal and transverse projectors with respect to momentum
\begin{equation}
    P_{ij}^L = \frac{p_i p_j}{\mathbf p^2}, \quad P_{ij}^T = \delta_{ij} - P_{ij}^L\,.
\end{equation}
We obtain
\begin{equation}
    K_{ij}(\mathbf p) = \mathbf p^2
    \left[
    \tilde\alpha P_{ij}^T + (\tilde\alpha + \beta) P_{ij}^L
    \right],
\end{equation}
which diagonalizes the kernel. As usual, we absorb the coefficient in front of the $\delta_{ij}$ term via a field redefinition, giving
\begin{equation}
    K_{ij}(\mathbf p) = \mathbf p^2
    \left[
    P_{ij}^T + (1 + h) P_{ij}^L
    \right],
    \quad h = \frac{\beta}{\tilde\alpha}.
\end{equation}
We can then see immediately that we require $h > -1$ for the theory to be stable. The full effective theory of the transition also includes local terms, encoding the mass parameter $m^2$ and self-interactions of the order parameter $\lambda$. In total, the real-space Euclidean action reads
\begin{equation}
    \mathcal S^\mathrm{cl} = \int_\mathbf{x} 
    \left[
    \frac 12 (\partial_i \phi_j)^2 + \frac h2 (\partial_i \phi_i)^2 + \frac 12 m^2 \phi_i^2 + \frac \lambda 4 (\phi_i^2)^2 
    \right]. \label{eq:Scl}
\end{equation}
We see that, physically, $h$ modifies the relative stiffness of longitudinal and transverse modes of the theory. At $h = 0$, both modes are equivalent and we recover $O(N)$ physics with $N = d$. Positive $h$ penalizes longitudinal fluctuations, whereas they are favored for negative $h$. At the stability boundary $h = -1$ the longitudinal stiffness vanishes, effectively suppressing transverse fluctuations. In contrast, longitudinal fluctuations are suppressed in the limit $h \to \infty$, where the theory becomes purely transverse, recovering the critical manifold of the long-range dipolar model~\cite{Aharony1973,Gimenez2024,Nakayama2024,Kalagov2026}.

\subsection{Quantum phase transition}
\label{sec:models-qpt}

At $T = 0$, we augment the order parameter with an additional dependence on imaginary time $\phi(\tau,\mathbf x)$. As such, the thermal action has to be supplemented with a term that encodes dynamics in imaginary time, giving
\begin{align}
    \mathcal S^\mathrm{q} = \int_{\tau, \mathbf{x}}
    \bigg[
    &\frac{1}{2 v_\phi^2} (\partial_\tau \phi_i)^2 + \frac 12 (\partial_i \phi_j)^2 + \frac h2 (\partial_i \phi_i)^2 \nonumber\\
    &+ \frac 12 m^2 \phi_i^2 + \lambda (\phi_i^2)^2 
    \bigg]. \label{eq:Squ}
\end{align}

Here, $v_\phi$ denotes the order parameter velocity, which can be set to unity by appropriate rescaling of $\tau$ and $\phi$. We emphasize that the quantum action Eq.~\eqref{eq:Squ} is not a na\"ive extension of the classical one Eq.~\eqref{eq:Scl} to higher dimension $d\rightarrow d+1$. The reason is that the imaginary time coordinate is not affected by spatial transformations so that only the trivial gradient term is symmetry-allowed and spatial and temporal contributions become non-equivalent. The longitudinal and transverse velocities are then given by 
\begin{equation}
    v_L = v_\phi \sqrt{1 + h}, \quad v_T = v_\phi,
\end{equation}
which differ at finite $h$, breaking Lorentz symmetry
\begin{equation}
    O(d+1) \to \mathcal T \times O(d).
\end{equation}
In the $h = 0$ subspace, Lorentz invariance is restored, which becomes apparent from $v_L = v_T$.

\subsection{Mode structure}

We start by discussing the structure of the eigenmodes of the classical action Eq.~\eqref{eq:Scl}. At tree level, these correspond to the eigenvalues of the Hessian of the microscopic action
\begin{equation}
    S_{ij}^{(2)}(\mathbf p) = \frac{\delta^2 S}{\delta \phi_i(-\mathbf p) \delta \phi_j(\mathbf p)} \bigg\vert_{\phi=\bar\phi},
\end{equation}
where $\bar\phi$ denotes an arbitrarily chosen vector in the manifold that minimizes $S$ and $\bar\phi_i$ is the i-th component of $\bar\phi$. In the symmetric phase with $m^2 > 0$ we find $S_{ij}^{(2)}(\mathbf p) = K_{ij}(\mathbf p) + m^2 \delta_{ij}$, corresponding to $d - 1$ transverse modes and a single longitudinal mode with eigenvalues
\begin{equation}
    \epsilon_T = \mathbf p^2 + m^2, \quad \epsilon_L = (1 + h) \mathbf p^2 + m^2.
\end{equation}
In the symmetry-broken phase with $m^2 < 0$, we have a finite minimum $\bar\phi^2 = -\tfrac{m^2}{\lambda}$ and thus
\begin{equation}
    S_{ij}^{(2)}(\mathbf p) = \mathbf p^2 \delta_{ij} + h p_i p_j + 2 \lambda \bar\phi_i \bar\phi_j.
\end{equation}
The structure of the eigenmodes becomes apparent if we introduce projectors with respect to the classical minimum of the action
\begin{equation}
    Q_{ij}^L = \frac{\bar\phi_i \bar\phi_j}{\bar\phi^2}, \quad Q_{ij}^T = \delta_{ij} - Q_{ij}^L.    
\end{equation}
With that, the Hessian becomes
\begin{align}
    S_{ij}^{(2)}(\mathbf p) = \mathbf p^2 \delta_{ij} + \mathbf p^2 h P_{ij}^L + m_R^2 Q_{ij}^L,
\end{align}
with the radial mass $m_R^2 = - 2 m^2$. We immediately see that the Hessian becomes block-diagonal, splitting into $d-2$ massless modes that are transverse with respect to momentum $\mathbf p$ and the condensate $\bar\phi$ with $\epsilon_T = \mathbf p^2$, which are the usual Goldstone modes, and a remaining two-dimensional longitudinal subspace, which we parameterize in the usual fashion, using an orthonormal basis oriented along the condensate. Projecting the full Hessian onto this basis gives
\begin{equation}
    \begin{bmatrix}
        \mathbf p^2 (1 + h \cos^2 \theta) + m_R^2  & \mathbf p^2 h \cos \theta \sin \theta \\
        \mathbf p^2 h \cos \theta \sin \theta & \mathbf p^2 (1 + h \sin^2 \theta)
    \end{bmatrix},
\end{equation}
where the angle $\theta$ parametrizes the orientation of the momentum relative to the condensate. To leading order in momentum, the two longitudinal eigenmodes read
\begin{align}
    \epsilon_- &= \mathbf p^2(1 + \sin^2 \theta) + \mathcal O(\mathbf p^4), \\
    \epsilon_+ &= m_R^2 + \mathbf p^2(1 + \cos^2 \theta) + \mathcal O(\mathbf p^4).
\end{align}
We thus find, in accordance with the Nambu-Goldstone theorem~\cite{Nambu1960,Goldstone1961,Goldstone1962}, that one of the two remaining modes is gapless, whereas the other one acquires the usual finite mass $m_R \propto \bar\phi$, meaning that in total we have $d - 1$ massless Goldstone modes and a single massive radial mode. However, the key difference to the $O(N)$ case is that the stiffnesses of the two modes in the longitudinal subspace are modulated by the angle of their momentum relative to the field expectation value. This is an explicit fingerprint of the spontaneously broken $O(d)$ rotation symmetry. The propagation of field excitations on a condensate background becomes direction-dependent.

Most of the discussion on the classical model carries over to the quantum model. More precisely, the eigenvalues of the Hessian of the classical model correspond to the on-shell frequencies of the quantum models, i.e., the physical poles of the real-time propagators. Thus, the phenomenology remains the same. In the symmetric phase, we obtain $d - 1$ transverse modes and a single longitudinal mode with velocities $v_T^2 = v_\phi^2$ and $v_L^2 = (1 + h) v_\phi^2$. In the symmetry-broken phase we observe a direction-dependent modulation of the velocities of the massive radial mode and one Goldstone mode, making the spontaneously broken rotational symmetry apparent.

Note, that although all couplings in the theory are renormalized in the IR, the general momentum structure of the theory is preserved under the RG flow, taken that we do not sit at the critical point. Therefore, away from criticality, the above findings remain valid in the IR. At the critical point, fluctuations induce a non-analytic momentum structure, which is captured by the presence of an RG fixed point.

\section{Renormalization Group approach}
\label{sec:RG}
\subsection{Scaling and critical exponents}

Critical physics is characterized by a divergence of the correlation length at the critical point associated with scale invariance of the system, which leads to the appearance of universal scaling laws for macroscopic observables. Physically, fluctuations on all length scales (and time scales, in case of quantum critical points) have to be accounted for to accurately describe the behavior of the system in the vicinity of the critical point. We do so via a RG approach~\cite{Kadanoff1966,Wilson1971,Wilson1971a} and successively integrate over high-energy modes, mapping the microscopic theory onto a renormalized one defined on a smaller energy scales. This procedure defines a flow in the space of couplings, which can be used to effectively interpolate between the microscopic theory defined at some ultraviolet (UV) scale $\Lambda$ and the full theory at the macroscopic level in the infrared (IR), taking all fluctuations into account.

In the RG picture, continuous phase transitions are described by fixed points of this flow in the space of couplings. To formalize this approach, we parametrize the theory space using a set of dimensionless couplings $\alpha_i$ with $i \in \{1, 2, \dots\}$. The flow  of these couplings is then described by the beta functions
\begin{equation}
    \partial_t \alpha_i = \beta_i(\alpha_1, \alpha_2, \dots),
\end{equation}
where $t = \ln \tfrac k\Lambda$ is the RG time, describing the change in scale $k$ relative to the UV scale $\Lambda$. More precisely, $t = 0$ at the UV scale where $k = \Lambda$ and $t \to -\infty$ at the IR scale where $k = 0$. Fixed points of the RG are then defined by the set of couplings $\alpha_i^\ast$ at which all beta functions vanish simultaneously
\begin{equation}
    \beta_i(\alpha_1^\ast, \alpha_2^\ast, \dots) = 0,
\end{equation}
since for this set of couplings $\partial_t \alpha_i = 0$ and the theory thus maps onto itself under the RG, rendering it scale invariant. The scaling behavior of the system near the transition emerges from the properties of the RG flow in the vicinity of the fixed point. We find the linearized flow
\begin{equation}
    \partial_t \alpha_i = M_{ij} (\alpha_j - \alpha_j^\ast) + \mathcal O[(\alpha_j - \alpha_j^\ast)^2],
\end{equation}
with the stability matrix
\begin{equation}
    M_{ij} = \frac{\partial \beta_i}{\partial \alpha_j} \bigg\vert_{\alpha = \alpha^\ast}.
\end{equation}
The critical exponents associated with the fixed points are the eigenvalues of $-M_{ij}$, which we denote by $\theta_i$ in the following. Positive $\theta_i > 0$ are RG relevant, meaning that the deviation from the fixed point in the direction of the respective eigenvector grows when flowing to the IR, whereas negative eigenvalues $\theta_i < 0$ are RG irrelevant and are attracted by the fixed point when flowing to the IR. The stability of fixed points can then be classified by the number of relevant couplings. While irrelevant couplings naturally flow into the fixed point in the IR, relevant couplings need to be fine-tuned to end up at the fixed point. This also reveals the physical interpretation of this picture. To access the phase transition, we have to tune a single parameter , temperature $T \to T_c$ in the thermal case or some non-thermal parameter $r \to r_c$ in the quantum case. In the RG picture, this amounts to a fixed point with a single relevant direction. This is what we call an IR stable fixed point or critical fixed point. Fixed points with more than one relevant direction need more fine-tuning to be physically accessible and are thus called multicritical fixed points.

To connect to experiments, it is convenient to introduce critical exponents that are associated with scaling laws of observables. The critical exponents are related to the eigenvalues of the stability matrix. In the theory literature, it is customary to quote the correlation length exponent $\nu = \theta_1^{-1}$, which describes the scaling of the correlation length
\begin{equation}
    \xi \propto \vert\Delta\vert^{-\nu},
\end{equation}
where $\Delta$ denotes the distance from the critical point in terms of some control parameter and $\theta_1$ is the largest relevant exponent. Furthermore, we calculate the anomalous dimension of the order parameter $\eta_\phi$, which describes the deviation of the scaling of the propagator from its canonical scaling
\begin{equation}
    G(\mathbf p) \propto \vert\mathbf p\vert^{-2 + \eta_\phi}.
\end{equation}
The remaining critical exponents can be related to those two through scaling and hyperscaling relations~\cite{Cardy1996}. Irrelevant eigenvalues modify the asymptotic scaling behavior, providing corrections to scaling that are quantified by their respective negative RG eigenvalue~\cite{Cardy1996}, e.g.
\begin{equation}
    \xi(\Delta) \sim A \vert \Delta \vert^{-\nu} (1 + B \vert \Delta \vert^{-\nu \omega} + \dots),
\end{equation}
where $A$ and $B$ are non-universal coefficients and $\omega = -\theta_2$ defines the leading correction-to-scaling exponent, given by the largest irrelevant RG eigenvalue. The dots denote subleading corrections associated with smaller irrelevant RG eigenvalues.

\subsection{Functional Renormalization Group}

\subsubsection{Method}

A modern implementation of Wilson's RG is the FRG approach~\cite{Wetterich1993,Ellwanger1994,Morris1994,Berges2002,Kopietz2010,Delamotte2012,Metzner2012,Dupuis2021}, whose formulation is based on the scale-dependent effective average action $\Gamma_k$. The latter interpolates between the microscopic (or bare) action $\Gamma_{k = \Lambda} = S$ and the full effective action $\Gamma_{k = 0} = \Gamma$, that is, the generating functional for one-particle-irreducible correlation functions. Within the FRG, its evolution is governed by an exact functional flow equation, the Wetterich equation~\cite{Wetterich1993}
\begin{equation}
    \partial_t \Gamma_k = \frac 12 \Tr 
    \left[
    \partial_t R_k 
    \left(
    \Gamma_k^{(2)} + R_k
    \right)
    ^{-1}
    \right],
\end{equation}
where the trace is over both spatial and internal coordinates. The right-hand side of the Wetterich equation includes the two-point function
\begin{equation}
    \Gamma_{k,ij}^{(2)}(p,q) = \frac{\delta^2 \Gamma_k}{\delta \phi_i(-p) \delta \phi_j(q)},
\end{equation}
and the infrared regulator $R_k$, which is formally introduced through a modification of the partition function
\begin{equation}
    \mathcal Z \to \mathcal Z_k = \int \mathcal D \phi \e^{-(S + \Delta S_k)},
\end{equation}
where the microscopic action is augmented by the regulator term
\begin{equation}
    \Delta S_k = \int_{p,q} \phi_i(-p) R_{k,ij}(p,q) \phi_j(q),
\end{equation}
making the partition function explicitly scale-dependent via the running cutoff scale $k$. Here, $p$ and $q$ denote general momentum variables, which, depending on the context, describe momenta for classical systems, and combined frequencies and momenta for quantum systems. The effective average action $\Gamma_k$ is then given by the Legendre transform of the scale-dependent Schwinger functional $W_k = \ln \mathcal Z_k$. Physically, the regulator acts as a scale-dependent mass, controlling the separation of fast and slow modes in the spirit of Wilson's momentum shell procedure. Fast modes with $p^2 \gg k^2$ are integrated, whereas the contribution of slow modes $p^2 \ll k^2$ to the partition function is suppressed. At the UV scale $k = \Lambda$, the regulator suppresses propagation of all modes, $R_{k = \Lambda} \to \infty$, in which case the effective action $\Gamma_k$ reduces to bare action, $\Gamma_{k = \Lambda} = S$. In the limit $k \to 0$ the regulator vanishes, $R_{k = 0} = 0$, restoring the original effective action with $\Gamma_{k = 0} = \Gamma$, such that all fluctuations enter the partition function.

\subsubsection{Derivative expansion}

Although the Wetterich equation is an exact equation, its solution generally requires some approximations, which are usually implemented by choosing a certain ansatz for $\Gamma_k$. The suitable approach to treat critical phenomena is the derivative expansion~\cite{Morris1994a,Morris1997,Morris1998,Morris1999,Berges2002,Balog2019,Morris2026}. The idea behind it is that the physics at the critical point is appropriately described by the long-wavelength behavior of correlation functions. Thus, it is sufficient to only keep the lowest terms in an expansion of the effective action in powers of momentum (or derivatives in real space) to obtain accurate estimates for the leading critical behavior. Such calculations have been performed successfully to high expansion orders on various different models~\cite{Canet2003,Kamikado2014,Heilmann2015,Knorr2016,Knorr2018,Divotgey2019,DePolsi2020,DePolsi2021,Knorr2021,Chlebicki2022,Sanchez2025}, including $\mathcal O(\partial^6)$ calculations for the Ising universality class~\cite{Balog2019}, showing excellent agreement with the most precise estimates from state-of-the-art numerical methods such as Monte Carlo simulations~\cite{Hasenbusch2021} and Conformal Bootstrap~\cite{Kos2014}.

The most general ansatz for the effective average action in the derivative expansion reads
\begin{equation}
    \Gamma_k = \int_x
    \left[
    U_k(\phi) + \sum_{n,\alpha} Z_{n,\alpha}(\phi) \mathcal O_{n,\alpha}
    \right],
\end{equation}
where $\mathcal O_{n,\alpha}$ denotes invariant operators at order $\mathcal O(\partial^n)$ of the derivative expansion and $\alpha$ enumerates all invariants within a given order. Here, $x$ denotes spatial coordinates in the classical case, but includes imaginary time in the quantum case. The $Z_{n,\alpha}(\phi)$ are the field-dependent coefficient functions for each operator. $U_k(\phi)$ denotes the effective potential, which encodes the full information of the theory at order $\mathcal O(\partial^0)$.

The $O(d) \times \mathds Z_2$ symmetry places some restrictions on the allowed values of $n$. Specifically, the $O(d)$ symmetry only permits operators $\propto \mathcal O(\partial^n \phi^m)$ with $n + m$ even, whereas the $\mathds Z_2$ symmetry dictates $m$ to be even. As a consequence, $n$ has to be even, giving $n \in \{2, 4, 6, \dots\}$. By symmetry, the field dependence of both the effective potential and the coefficient function $Z_{n,\alpha}$ enters only through the field invariant
\begin{equation}
    \bar\rho = \frac 12 \phi^2_i.
\end{equation}
Flow equations for $U_k(\bar\rho)$ and $Z_{n,\alpha}(\bar\rho)$ can be derived by projecting the Wetterich equation onto the respective invariant operator $\mathcal O_{n,\alpha}$. A blueprint for the derivation of projectors to arbitrary order in the derivative expansion is given in App.~\ref{app:proj-andim}.

\section{FRG flow equations}
\label{sec:FRG}
\subsection{Thermal phase transition}

\subsubsection{Effective average action}

We start by deriving flow equations for the classical action. Although a systematic study of several orders in the derivative expansion has proven itself to be well-suited for obtaining precise estimates for critical exponents and other universal quantities, it usually requires considerable numerical effort. Here, we aim for a different strategy, employing instead a truncation that allows us to carry out most calculations analytically. This truncation is known as the extended LPA or LPA'~\cite{Morris1994a,Adams1995,Berges1996,Berges2002,Blaizot2006} and considers, beyond the effective potential, only the flow of the leading kinetic terms. As such, it can be interpreted as a partial $\mathcal O(\partial^2)$ truncation. In addition, we approximate the field-dependent renormalization factors of the leading kinetic terms by their value at the minimum of the effective potential, giving the ansatz
\begin{align}
    \Gamma_k &= \int_\mathbf{x}
    \left[
    U_k(\bar\rho) + \frac 12 Z_\phi (\partial_i \phi_j)^2 + \frac 12 Z_h (\partial_i \phi_i)^2
    \right] \\
    &= \int_\mathbf{x}
    \left\{
    U_k(\bar\rho) + \frac 12 Z_\phi 
    \left[
    (\partial_i \phi_j)^2 + h_k (\partial_i \phi_i)^2
    \right]
    \right\},
\end{align}
where we have introduced an anisotropy factor given by the ratio of the two renormalization factors
\begin{equation}
    h_k = \frac{Z_h}{Z_\phi}.
\end{equation}
Our LPA' ansatz presents the minimal truncation that is able to describe both non-Gaussian critical correlations, enabled by a non-trivial field renormalization of $Z_\phi$, and separate longitudinal and transverse stiffnesses
\begin{equation}
    \sigma_L = 1 + h_k, \quad \sigma_T = 1.
\end{equation}
As such, we expect this ansatz to be sufficient to capture the essential physics of the model at hand.

\subsubsection{Regulators}

Whilst any physical observables obtained from the Wetterich equation are in principle independent of the choice of regulator, truncations will generically introduce a regulator dependence to the flow equations in any practical calculation. Because of that, we aim to employ a suitable regulator that minimizes this dependence. A general guiding principle is to use a regulator that respects the symmetries of the theory\footnote{Technically, this is necessary to ensure that the Ward-Takahashi identities of the full effective action are satisfied at arbitrary cutoff scales $k$.}. This is especially important in the context of the study of emergent symmetries since it avoids spurious regulator-induced symmetry breaking during the flow. 

For our problem, for example, the microscopic action reduces to the $O(N)$ model for $h = 0$. As such, we need to employ a regulator that respects the $O(N)$ symmetry in this limit, because otherwise the flow will leave the $O(N)$ subspace since the flow equations themselves do not adhere to the symmetry due to the residual symmetry breaking through the regulator.

These considerations leave us with two possible regulators. First, we consider a symmetric regulator that is independent of $h_k$, and thus fully invariant under the $O(N)$ symmetries
\begin{equation}\label{eq:sym-reg}
    R_{k,ij}^\mathrm{sym}(\mathbf p,\mathbf q) = Z_\phi \mathbf p^2 \delta_{ij} r_k(\mathbf p^2/k^2) \delta(\mathbf p-\mathbf q).
\end{equation}
For the shape function, we introduce the linear regulator, which has been shown to minimize the regulator dependence in $O(N)$ models at the level of the LPA~\cite{Litim2001,Litim2002,Canet2003a}
\begin{equation} \label{eq:shapefct}
    r_k(x) =
    \left(
    \frac 1x - 1
    \right)
    \Theta(1-x),
\end{equation}
which suppresses slow modes with $\mathbf p^2 \ll k^2$ only gradually. Although the symmetric regulator recovers all necessary symmetries, it leaves room for improvement. At finite $h_k$, the regulator affects both longitudinal and transverse modes in the same way, even though their stiffnesses differ through $h_k$. As a consequence, the phase space volume of modes suppressed by the regulator is different for longitudinal and transverse modes. To rectify this, we introduce a phase-space adjusted regulator~\cite{Pawlowski2007}
\begin{equation}\label{eq:adj-reg}
    R_{k,ij}^\mathrm{adj}(\mathbf p,\mathbf q) = Z_\phi (\mathbf p^2 \delta_{ij} + h_k p_i p_j) r_k(\mathbf p^2/k^2) \delta(\mathbf p-\mathbf q),
\end{equation}
which now explicitly depends on $h_k$. We note that the presence of an additional scale-dependent quantity $h_k$ in the regulator will produce an additional term proportional to $h_k$ in any flow equation
coming from the scale-derivative of the regulator itself
\begin{align}
    \partial_t R_{k,ij}^\mathrm{adj} &= Z_\phi
    (\partial_t r_k - \eta_\phi r_k) \mathbf p^2 \delta_{ij} \nonumber\\
    &\phantom{=}
    + Z_\phi (\partial_t r_k - \eta_h r_k) h_k p_i p_j.
\end{align}
In combination with the linear shape function~\eqref{eq:shapefct}, this regulator allows us to fully resolve the radial integrals appearing in the FRG loop integrals, simplifying the resulting expressions.

\subsubsection{Flow equations}

In the following, we aim to derive flow equations for the scale-dependent quantities introduced in our ansatz. To make the scaling behavior of fixed-point solutions to the flow equations apparent, it is convenient to introduce the dimensionless effective potential
\begin{equation}
    u(\rho) = k^{-d} U_k(\bar\rho),
\end{equation}
and to express all flow equations in terms of the dimensionless renormalized field invariant $\rho = Z_\phi k^{2-d} \bar\rho$. Additionally, we introduce anomalous dimensions
\begin{equation}
    \eta_\phi = -\partial_t \ln Z_\phi, \quad \eta_h = -\partial_t \ln Z_h,
\end{equation}
with which the scaling dimension of the field becomes
\begin{equation}
    \Delta_\phi = \frac{d - 2 + \eta_\phi}{2}.
\end{equation}
We obtain the flow equation for the dimensionful effective potential $U_k(\bar\rho)$ by evaluating the Wetterich equation at a constant field configuration $\bar\rho(x) = \mathrm{const.}$, which corresponds to setting $\phi(p) = \phi\, \delta(p)$ and thus gives
\begin{equation}
    \partial_t U_k(\bar\rho) = \frac 1\Omega \partial_t\Gamma_k \big\vert_{\bar\rho = \mathrm{const.}},
\end{equation}
where $\Omega$ denotes the volume. For the dimensionless effective potential, we thus obtain
\begin{align}\label{eq:flowu}
    \partial_t u(\rho) =& -d u(\rho) + (d - 2 + \eta_\phi) \rho u'(\rho) \nonumber\\
    &+ (d - 2) I_G(u'(\rho), \eta_\phi)  \nonumber\\
    &+ I_R(u'(\rho), 2\rho u''(\rho), h_k, \eta_\phi) \nonumber\\
    &+ I_h(u'(\rho), 2\rho u''(\rho), h_k, \eta_\phi,\eta_h).
\end{align}
Here, the first line denotes the contributions due to the intrinsic scaling of the dimensionful potential, whereas the remaining terms are fluctuation contributions. The threshold function $I_G$ encodes the contributions of the $d - 2$ Goldstone modes that are transverse with respect to both external momentum and condensate, while $I_R$ describes the contribution of the two longitudinal modes. In the $h_k = 0$ subspace, this term cleanly separates into Goldstone and radial contributions, recovering the flow equation for the $O(N)$ model with $N = d$~\cite{Berges2002}. For the phase-space adjusted regulator, we also obtain an additional contribution through the scale derivative of the regulator that is encoded in the threshold function $I_h$. The expressions for the threshold functions are given in App.~\ref{app:thresholds}.

For the identification of RG fixed points and the calculation of their associated critical exponents, it is convenient to expand the dimensionless effective potential about its scale-dependent minimum $\kappa$. For the present problem we find that the fixed point lies in the symmetry-broken regime with finite minimum $\kappa \neq 0$
\begin{equation}\label{eq:pot-exp}
    u(\rho) = \sum_{n = 2}^\infty \frac{\lambda_n}{n!} (\rho - \kappa)^n,
    \quad
    \lambda_n = \frac{\partial^n}{\partial \rho^n} u(\rho) \bigg\vert_{\rho = \kappa}.
\end{equation}
We have thus traded the flow equation for the dimensionless potential, which is a partial differential equation, for an infinite tower of ordinary differential equations for the expansion coefficients $\lambda_n$ and $\kappa$, which, in practice, we truncate at $n=8$ where our numerical results have already converged. We obtain flow equations for the coefficients by appropriately projecting Eq.~\eqref{eq:flowu} onto the expansion, giving
\begin{align}
    \partial_t \lambda_n &= \lambda_{n+1} \partial_t \kappa + \frac{\partial^n}{\partial \rho^n} (\partial_t u) \bigg\vert_{\rho = \kappa}, \label{eq:flow-lambda}\\
    \partial_t \kappa &= -\frac{1}{\lambda_2} \frac{\partial}{\partial \rho} (\partial_t u) \bigg\vert_{\rho = \kappa}. \label{eq:flow-kappa}
\end{align}
Expressions for the anomalous dimensions are derived from the flow of the renormalization factors $Z_\phi$ and $Z_h$, which are obtained by projecting on the appropriate tensor structures of the two-point function at constant field
\begin{equation}
    \left(
    \Gamma_{k,0}^{(2)} 
    \right)_{ij}
    = 
    \left[
    U_k^\prime(\rho) + Z_\phi \mathbf p^2
    \right]
    \delta_{ij} + Z_h p_i p_j + U_k^{\prime\prime}(\rho) \phi_i \phi_j.
\end{equation}
From there, we obtain the projections
\begin{align}
    \partial_t Z_\phi &= \Pi_{ab,ij}^\phi \frac{\partial^2 }{\partial p_a \partial p_b} \partial_t
    \left(
    \Gamma_{k,0}^{(2)} 
    \right)_{ij}
    \bigg\vert_{\substack{\mathbf p = 0 \\ \bar\rho = \bar\kappa}}, \\
    \partial_t Z_h &= \Pi_{ab,ij}^h \frac{\partial^2 }{\partial p_a \partial p_b} \partial_t
    \left(
    \Gamma_{k,0}^{(2)} 
    \right)_{ij}
    \bigg\vert_{\substack{\mathbf p = 0 \\ \bar\rho = \bar\kappa}}.
\end{align}
Here, $\bar\kappa$ denotes the value of the field invariant at the minimum of the dimensionful potential. The renormalization factors capture the flow of $h_k$, which is conveniently expressed in terms of the anomalous dimensions
\begin{equation} \label{eq:beta_h1}
    \partial_t h_k = \partial_t
    \left(
    \frac{Z_h}{Z_\phi}
    \right)
    = h_k (\eta_\phi - \eta_h).
\end{equation}
The detailed calculation of the anomalous dimensions and the flow equation of $h_k$ is given in App.~\ref{app:flow-andim}. The derivation of the projection operators $\Pi^\phi$ and $\Pi^h$ can be found in App.~\ref{app:proj-andim}.

Note that the projection for the anomalous dimension is different from the one that is usually used in the treatment of $O(N)$ models. There, it is customary to extract the flow of $Z_\phi$ by projecting onto the Goldstone modes, or, more generally, onto modes perpendicular to the minimum of the effective potential to avoid contamination of the flow from higher-order terms in the derivative expansion that have been neglected in the LPA' truncation~\cite{Wetterich1993a,Tetradis1994}. A na\"ive extension of this procedure to the present $O(d)$ models fails due to the enlarged tensor basis of the two-point function, leading to mixing of the flow of $Z_\phi$ and $Z_h$. A detailed description of this phenomenon is given in App.~\ref{app:proj-andim}.

\subsection{Quantum phase transition}

\subsubsection{Effective average action}

For the quantum model, the situation changes only slightly. As in the classical case, we employ an LPA' truncation with constant renormalization factors. As discussed in Sec.~\ref{sec:models-qpt}, this amounts to adding a dynamical term to the classical model, giving
\begin{align}
    \Gamma_k &= \int_{\tau,\mathbf{x}}
    \bigg[
    U_k(\bar\rho) + \frac 12 Z_{\phi,\parallel} (\partial_\tau \phi_j)^2 + \frac 12 Z_{\phi,\perp} (\partial_i \phi_j)^2 \nonumber\\
    &\phantom{= \int_x \bigg[ U_k(\bar\rho)}
    + \frac 12 Z_h (\partial_i \phi_i)^2
    \bigg] \\
    &= \int_{\tau,\mathbf{x}}
    \Bigg\{
    U_k(\bar\rho) + \frac 12 Z_{\phi,\perp}
    \Bigg[
    \frac{1}{v_{\phi,k}^2} (\partial_\tau \phi_j)^2 + (\partial_i \phi_j)^2\nonumber\\
    &\phantom{= \int_x \bigg[ U_k(\bar\rho)}
    + h_k (\partial_i \phi_i)^2
    \Bigg]
    \Bigg\}, \label{eq:queffaction}
\end{align}
where we have introduced the scale-dependent velocity and the anisotropy factor
\begin{equation}
    v_{\phi,k} = 
    \left(
    \frac{Z_{\phi,\perp}}{Z_{\phi,\parallel}}
    \right)
    ^{1/2}, \quad h_k = \frac{Z_h}{Z_{\phi,\perp}}.
\end{equation}
Note that there is one more flowing coupling in the quantum case. As before, the general theory breaks the symmetry between longitudinal and transverse modes. In the quantum theory, this is expressed in terms of the respective velocities
\begin{equation}
    v_L = v_{\phi,k} \sqrt{1 + h_k}, \quad v_T = v_{\phi,k},
\end{equation}
with $v_L \neq v_T$ signaling a broken Lorentz symmetry. However, even for matching velocities the presence of independent renormalization factors $Z_{\phi,\parallel}$ and $Z_{\phi,\perp}$ explicitly breaks Lorentz symmetry. Their relative scaling at the QCP defines the dynamical critical exponent $z$. With $[\mathbf p] = 1$, it is defined as the scaling dimension of frequencies $[\omega] = z$. Then, the scaling dimension of the velocity becomes $[v_{\phi,k}] = z - 1$ and we find
\begin{equation}
    z = 1 + \partial_t \ln v_{\phi,k} = 1 + \frac{\eta_{\phi,\parallel} - \eta_{\phi,\perp}}{2},
\end{equation}
with anomalous dimensions
\begin{equation}
    \eta_{\phi,\parallel} = -\partial_t \ln Z_{\phi,\parallel}, \quad \eta_{\phi,\perp} = -\partial_t \ln Z_{\phi,\perp}.
\end{equation}
Hence, Lorentz invariance is only restored if $h_k = 0$ and $z = 1$, corresponding to matching longitudinal and transverse velocities $v_L = v_T = v_{\phi,k}$ and trivial scaling $\partial_t v_{\phi,k} = 0$. The scaling dimension of the field becomes
\begin{equation}
    \Delta_\phi = \frac{d + z - 2 + \eta_{\phi,\perp}}{2}.
\end{equation}

\subsubsection{Regulators}

For the choice of regulator, we are faced with a situation similar to that of the thermal model. In the presence of quantum fluctuations, we have the choice between two types of regulators that preserve the symmetries of the UV theory during the flow. First, a manifestly covariant regulator
\begin{equation}\label{eq:cov-regQ}
    R_{k,ij}^\mathrm{cov}(p,q) = Z_{\phi,\perp} p^2 \delta_{ij} r_k(p^2/k^2) \delta(p-q),
\end{equation}
and second, a spectrally adjusted regulator
\begin{equation}\label{eq:adj-regQ}
    R_{k,ij}^\mathrm{adj}(p,q) = Z_{\phi,\perp} (p^2 \delta_{ij} + h_k p_i p_j) r_k(p^2/k^2) \delta(p-q),
\end{equation}
which takes into account the difference in spectral density between longitudinal and transverse fluctuations, but restores Lorentz symmetry in the $h_k = 0$ subspace, as required. For both regulators we have introduced
\begin{equation}
    p^2 = \frac{\omega^2}{v_{\phi,k}^2} + \mathbf p^2.
\end{equation}
We use the same linear shape function defined in Eq.~\eqref{eq:shapefct} as in the classical case.

\subsubsection{Flow equations}

We again consider the flow equation for the effective potential first. Due to the presence of the additional imaginary-time integral in the action Eq~\eqref{eq:queffaction}, we need to adapt the definition of dimensionless renormalized quantities to the quantum case, giving
\begin{equation}
    u(\rho) = v_{\phi,k}^{-1} k^{-d-1} U_k(\bar\rho),
    \quad
    \rho = Z_{\phi,\perp} v_{\phi,k}^{-1} k^{1-d} \bar\rho.
\end{equation}
We then obtain the flow equation for the dimensionless renormalized effective potential
\begin{align}
    \partial_t u(\rho) =& -(d + z) u(\rho) + (d + z - 2 + \eta_{\phi,\perp}) \rho u'(\rho) \nonumber\\
    &+ (d - 2) I_G(u'(\rho), \eta_{\phi,\perp},z)  \nonumber\\
    &+ I_R(u'(\rho), u'(\rho), 2\rho u''(\rho), h_k, \eta_{\phi,\perp},z) \nonumber\\
    &+ I_h(u'(\rho), u'(\rho), 2\rho u''(\rho), h_k, \eta_{\phi,\perp},z),
\end{align}
with the threshold functions again given in App.~\ref{app:thresholds}. We observe that the key difference is the appearance of the dynamical critical exponent in the dimensional contribution to the flow and as an additional argument to the threshold functions. Otherwise, the structure of the fluctuation contributions is equivalent to what we found for the thermal model. Similarly, the field expansion of the potential also carries over from the thermal case.

Starting point for the derivation of the anomalous dimensions $\eta_{\phi,\parallel}$, $\eta_{\phi,\perp}$ and $\eta_h$ is again the two-point function
\begin{align}
    \left(
    \Gamma_{k,0}^{(2)} 
    \right)_{ij} &= 
    \left[
    U_k^\prime(\rho) + Z_{\phi,\parallel} \omega^2 + Z_{\phi,\perp} \mathbf p^2
    \right]
    \delta_{ij} \nonumber\\
    &\phantom{=}+ Z_h p_i p_j + U_k^{\prime\prime}(\rho) \phi_i \phi_j.
\end{align}
We observe that terms $\mathcal O(\omega \mathbf p)$, mixing spatial and temporal sectors, are forbidden by symmetry because frequencies and spatial momenta transform under different irreducible representations of the symmetry group. Due to the absence of mixing terms in the two-point function, the projectors for the spatial renormalization factors $Z_{\phi,\perp}$ and $Z_h$ remain the same as in the classical case
\begin{align}
    \partial_t Z_{\phi,\perp} &= \Pi_{ab,ij}^\phi \frac{\partial^2 }{\partial p_a \partial p_b} \partial_t\Gamma_{k,ij}^{(2)} \Bigg\vert_{\substack{\omega = \mathbf p = 0 \\ \bar\rho = \bar\kappa}}, \\
    \partial_t Z_h &= \Pi_{ab,ij}^h \frac{\partial^2 }{\partial p_a \partial p_b} \partial_t\Gamma_{k,ij}^{(2)} \Bigg\vert_{\substack{\omega = \mathbf p = 0 \\ \bar\rho = \bar\kappa}}.
\end{align}
As before, we find for the flow of $h_k$
\begin{equation}\label{eq:beta_h2}
    \partial_t h_k = h_k (\eta_{\phi,\perp} - \eta_h).
\end{equation}
The flow of the temporal renormalization factor $Z_{\phi,\parallel}$, which is necessary for the calculation of the dynamical critical exponent, is obtained by projecting onto the frequency dependence of the two-point function
\begin{equation}
    \partial_t Z_{\phi,\parallel} = P_{ij}^T(\phi) \frac{\partial^2 }{\partial \omega^2} \partial_t\Gamma_{k,ij}^{(2)} \Bigg\vert_{\substack{\omega = \mathbf p = 0 \\ \bar\rho = \bar\kappa}}.
\end{equation}
A detailed derivation of the projectors and the resulting flow equations is given in App.~\ref{app:proj-andim}.

\section{Critical behavior}
\label{sec:CriticalBehavior}

\begin{table}
    \centering
    \caption{Critical exponents $\nu$ and $\eta_\phi$ and the correction-to-scaling exponent $\omega_h$ for the classical $O(d)$ model in 3 dimensions and the quantum $O(d)$ model in 2+1 dimensions. We compare with results from complementary approaches, including higher-order FRG truncations, perturbative RG, and results from conformal bootstrap (CB) and Monte Carlo simulations (MC).}
    \begin{tabular*}{\columnwidth}{@{\extracolsep{\fill}}llll}
        \hline\hline \\
        3-dim. classical $O(d)$ & \multicolumn{1}{c}{$\nu$} & \multicolumn{1}{c}{$\eta_\phi$} & \multicolumn{1}{c}{$\omega_h$} \\
        \hline \\
        FRG, LPA16' (\textit{this work}) & 0.7318 & 0.0409 & 0.0330 \\
        FRG, $\mathcal O(\partial^2)$~\cite{DePolsi2020} & 0.7125(71) & 0.0408(58) & \\
        FRG, $\mathcal O(\partial^4)$~\cite{DePolsi2020}
        & 0.7114(9) & 0.0376(13) & \\
        $\epsilon^2$~\cite{Aharony1973a} & 0.6875 & 0.0208 & 0.0092\\
        $\epsilon^4$~\cite{Adzhemyan2023} & & & 0.0156(3) \\
        $\epsilon^6$~\cite{Kompaniets2017} & 0.7059(20) & 0.0378(5) & \\
        MC~\cite{Hasenbusch2011} & 0.7116(10) & 0.0378(3) & \\
        CB~\cite{Kos2016} & 0.7120(23) & 0.0385(13) & \\
        \hline\\
        2+1-dim. quantum $O(d)$ & \multicolumn{1}{c}{$\nu$} & \multicolumn{1}{c}{$\eta_\phi$} & \multicolumn{1}{c}{$\omega_h$} \\
        \hline \\
        FRG, LPA16' (\textit{this work}) & 0.6860 & 0.0437 & 0.0353 \\
        FRG, $\mathcal O(\partial^2)$~\cite{DePolsi2020} & 0.6725(52) & 0.0410(59) & \\
        FRG, $\mathcal O(\partial^4)$~\cite{DePolsi2020} & 0.6716(6) & 0.0380(13) & \\
        $\epsilon^6$~\cite{Kompaniets2017} & 0.6690(10) & 0.0380(6) & \\
        MC~\cite{Hasenbusch2025} & 0.671718(23) & 0.03816(2) & \\
        CB~\cite{Chester2020} & 0.6718(1) & 0.03818(4) & \\
        \hline\hline
    \end{tabular*}
    \label{tab:exponents}
\end{table}

\begin{figure*}[tp]
    \centering
    \includegraphics[width=\textwidth]{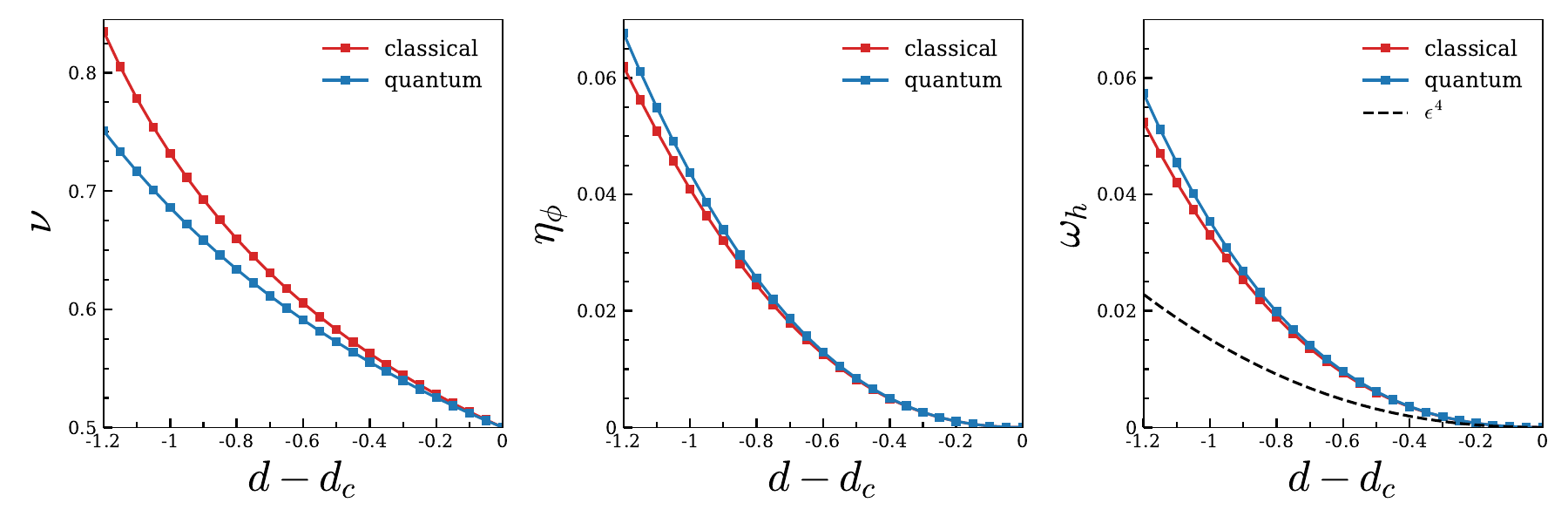}
    \caption{Correlation length exponent $\nu$ (left panel), anomalous dimension $\eta_\phi$ (center panel) and leading correction-to-scaling exponent $\omega_h$ (right panel) for the classical (red) and quantum (blue) $O(d)$ universality classes as a function of spatial dimension. The horizontal axis denotes the distance from the upper critical dimension $d_c$, with $d_c = 4$ ($d_c = 3$) in the classical (quantum) model. The results from the $\epsilon$-expansion shown on the right panel are taken from Ref.~\cite{Adzhemyan2023}.}
    \label{fig:exponents-dim}
\end{figure*}

\subsection{\texorpdfstring{$h_k = 0$}{h\_k = 0} fixed points}

From the structure of the flow equation for $h_k$, see Eqs.~\eqref{eq:beta_h1},~\eqref{eq:beta_h2}, we immediately see that $h_k = 0$ is a fixed point of the RG flow. This is expected because $h_k = 0$ presents a high-symmetry subspace of the theory space, where the usual $O(N)$ symmetries are restored, rendering both thermal and quantum models invariant under separate rotations of field and momenta. In this subspace, classical and quantum models are described by the same set of equations. To be precise, the classical (quantum) model reduces to the $O(N=d)$ model in $d$ ($d+1$) dimensions. Note, however, that the dimensional dependence of the resulting fixed points does not coincide with the $O(N)$ results, since the number of field components is strictly tied to the spatial dimension of the system at hand. Explicitly, this means that we interpolate from $O(4)$ to $O(3)$ for $3\leq d\leq 4$ in the classical case and from $O(3)$ to $O(2)$ for $2\leq d\leq 3$ in the quantum case.

In the following, we analyze the fixed-point structure of both classical and quantum models in the $h_k = 0$ subspace and show that adding the coupling $h_k$ leads to an irrelevant perturbation. We recover the usual critical behavior described by the IR stable Wilson-Fisher fixed point below the upper critical dimension of $d_c = 4$ ($d_c = 3$) in the classical (quantum) case by using the field expansion of the effective potential introduced in Eqs.~\eqref{eq:pot-exp}--\eqref{eq:flow-kappa}. Additionally, we calculate an FRG estimate for the correction-to-scaling exponent due to $h_k$. To establish convergence of our results with respect to the potential expansion, we truncate the expansion at increasing orders $n \in \{2,3,4,5,6,7,8\}$, corresponding to $2n$ orders in the field.

We quote our results for the correlation length exponent $\nu$ and the anomalous dimension $\eta_\phi$ for both the classical model in $d = 3$ and the quantum model in $d = 2$ in Table~\ref{tab:exponents}. These exponents were already calculated from an LPA' truncation within FRG for the $O(N=3)$ model, in particular in Refs.~\cite{Adams1995,Berges1996} where an exponential regulator was used. The quick convergence of the derivative expansion was demonstrated in Ref.~\cite{Balog2019}. Thus, we also find that our results show good numerical agreement with other methods, including higher-order truncations of the derivative expansion~\cite{DePolsi2020} and numerically exact results from conformal bootstrap~\cite{Kos2016,Chester2020} and Monte Carlo simulations~\cite{Hasenbusch2011,Hasenbusch2025}, especially in comparison to low-order perturbative calculations, see e.g. Refs.~\cite{Aharony1973a} and~\cite{Herbut2007}. The values of the critical exponents in arbitrary dimensions are shown in the left and center panels of Fig.~\ref{fig:exponents-dim}.

\subsection{Stability and correction to scaling}

To assess the stability of the Wilson-Fisher fixed points against perturbations out of the $h_k = 0$ subspace, we need to consider the full flow equations with finite $h_k$. Most importantly, we now include the beta function for $h_k$ itself. To calculate the RG eigenvalue associated with finite $h_k$ perturbations, we make use of the fact that $h_k = 0$ is an invariant subspace of the RG flow. Because $\beta_h$ vanishes identically in this invariant subspace, variations of the $O(N)$ couplings arising from the potential expansion, which reside in this subspace, cannot generate a flow out of this subspace
\begin{equation}
    \frac{\partial \beta_h}{\partial \kappa} \Bigg\vert_{h_k = 0} = \frac{\partial \beta_h}{\partial \lambda_i} \Bigg\vert_{h_k = 0} = 0.
\end{equation}
Hence, the stability matrix takes a block-triangular form that separates RG eigenvalues of the invariant $h_k = 0$ subspace and its complement. Consequently, the RG eigenvalue associated with $h_k$ at the Wilson-Fisher fixed points is given by
\begin{equation}\label{eq:scaling-exponent}
    y_h = -\frac{\partial \beta_h}{\partial h_k} \Bigg\vert_{h_k = 0} =
    \begin{cases}
        \eta_h - \eta_\phi & \mathrm{classical} \\
        \eta_h - \eta_{\phi,\perp} & \mathrm{quantum}
    \end{cases}.
\end{equation}
Concrete evaluation provides analytic expressions for the exponents
\begin{equation}
    y_h^\mathrm{cl} = -\frac{4 + \eta_\phi}{d + 2} \eta_\phi,
    \quad
    y_h^\mathrm{q} = -\frac{4 + \eta_{\phi,\perp}}{d + 3} \eta_{\phi,\perp}
\end{equation}
and therefore depends only on the spatial dimension $d$ and the anomalous dimension at the Wilson-Fisher fixed point. In particular, we observe that because the anomalous dimension is strictly positive below the upper critical dimension, as is evident from Fig.~\ref{fig:exponents-dim}, $y_h$ is always negative, confirming the results of Refs.~\cite{Aharony1973a} and~\cite{Adzhemyan2023}. Hence, $h_k$ is irrelevant at the Wilson-Fisher fixed point, which therefore describes the asymptotic critical behavior of the $O(d)$ system. Consequently, we find an emergent $O(N=d)$ symmetry at the critical point for both classical and quantum systems. 

We also see that the quantum-to-classical correspondence $d\rightarrow d+1$ holds for $y_h$, which is expected for the critical surface of the standard $O(N)$ model $h_k=0$, but not necessarily for the exponent associated to its perturbation since this is not a general feature of the RG equations. This becomes apparent, for example, by expanding the threshold functions for the effective potential in a polynomial series in $h_k$. Evaluating the series coefficients order by order reveals that the quantum-to-classical mapping holds only at leading order, which corresponds to the $O(N)$-invariant subspace. Interestingly, however, numerical evaluation of the flow equations at random couplings suggests that $\eta_{\phi,\parallel} = \eta_{\phi,\perp}$, and thus $z = 1$, is a general feature of the quantum model. Furthermore, we note that the expressions for the anomalous dimensions, and thus for $y_h$, are independent of the regulator used. The regulator dependence only enters non-linear contributions to the flow in the vicinity of the fixed point. 

From Eq.~\eqref{eq:scaling-exponent}, as well as from the right panel of Fig.~\ref{fig:exponents-dim}, we find that $y_h$ is of the order of the anomalous dimension. Although this does not influence the asymptotic scaling at the critical point, it has a profound influence on the crossover between the non-universal regime that is dominated by short-range fluctuations and the asymptotic scaling regime. Here, irrelevant couplings have not died out yet, providing corrections to scaling. In the theory without $h_k$, we find the correction-to-scaling exponent  
to be relatively large. Within our LPA' calculation, we obtain for the classical (quantum) model in $d = 3$ ($d = 2$) a value of $\omega = 0.7500$ ($\omega = 0.7348$). Consequently, in the standard $O(N)$ corrections to scaling approach zero at small length scales so that we can expect the appearance of an extended asymptotic scaling regime. The situation is very different if the UV theory has a finite $h_k$. Due to its weakly irrelevant nature, $h_k$ decays very slowly towards the IR, and only vanishes at larger scales. Hence, asymptotic critical behavior is only expected to show up very close to the critical point, with a large region of considerable corrections to scaling, assuming no fine-tuning of $h_k$ in the UV. The associated correction-to-scaling exponent is
\begin{equation}
    \omega_h = -y_h.
\end{equation}
Numerical estimates for both classical and quantum models are provided in Table \ref{tab:exponents}. Its dimensional dependence is displayed in the right panel in Fig.~\ref{fig:exponents-dim}. In principle, the presence of these much more pronounced corrections to scaling should show up in both Monte Carlo simulations and experiments. Finally, we note that the relative difference between our estimate for $\omega_h$ for the classical model and the one obtained from high-order $\epsilon$-expansion~\cite{Adzhemyan2023} is large compared to the relative difference in the other critical exponents, $\Delta \omega_h /\omega_h \simeq 50\%$. To compare the results between methods, we need to consider the relative difference of fluctuation corrections obtained from these methods. For the established critical exponents, we find a margin of $\Delta\nu / \nu \simeq 9 \dots 11\%$ for the correlation length exponent and $\Delta \eta_\phi /\eta_\phi \simeq 7 \dots 10\%$ for the anomalous dimension, when compared against the $\epsilon^6$ results~\cite{Kompaniets2017}. Comparison of our estimates for the $O(N)$ correction-to-scaling exponent with results from more precise truncations and methods, see e.g. Ref.~\cite{DePolsi2020}, suggest that the discrepancy cannot be attributed to the fact that $\omega_h$ is an irrelevant eigenvalue and thus requires a different explanation.

\section{Conclusions}
\label{sec:Conclusions}

In this work, we have studied the critical behavior of a class of structural phase transitions in isotropic media, which are driven by fluctuations of non-polar order parameters. Non-polar orders are, in addition to their non-trivial transformation properties under spatial $O(d)$ rotations, characterized by the presence of an additional internal $\mathds Z_2$ symmetry. In physical terms, this internal $\mathds Z_2$ symmetry is realized in orders that are odd under time reversal, including antiferromagnets and spin-orbit-coupled magnets, but also in inversion-even orders like ferroaxial and ferrotoroidic systems.~\cite{Aharony1973,Aharony1973a,Aharony1973b,Bak1980,Frey1994,Gnewuch2019,Picozzi2024,Salje1990,Salje2005,Hlinka2016,Fernandes2019,Binghai2024}

In an effective field theory, these systems are described by a theory with interlocked internal and spatial $O(d)$ symmetry, augmented by an additional internal $\mathds Z_2$ symmetry. This interlocking translates into an extra kinetic term proportional to a coupling $h$, which can be viewed as a perturbation to the conventional $O(N)$ models. We have derived FRG equations within the LPA' for both the classical and quantum versions of this $O(d)$ model, taking into account the enlarged tensor structure of the two-point function, which is essential for consistently disentangling the different kinetic couplings. Our fixed-point analysis reveals that the additional coupling $h$ is irrelevant at the Wilson-Fisher fixed point in both cases, leading to an emergent $O(N=d)$ symmetry at criticality. The resulting estimates for the correlation-length exponent and anomalous dimension are in good overall agreement with results from higher-order FRG calculations, perturbative RG, Monte Carlo simulations, and conformal bootstrap.

We further find that the RG eigenvalue associated with $h$ is negative and can be expressed analytically in terms of the anomalous dimension at the fixed point. This extends the previously established result for classical transitions~\cite{Aharony1973a,Adzhemyan2023} to the quantum case. We also investigated the relation between classical and quantum critical behaviors. The conventional quantum-to-classical mapping, $d\rightarrow d+1$, is preserved on the $h=0$ subspace (i.e. the $O(N)$-symmetric subspace), but does not generally extend to the full RG flow at finite $h$. Nevertheless, our numerical results suggest that the temporal and spatial anomalous dimensions remain equal within the quantum model, implying $z=1$ even away from the $O(N)$-symmetric subspace. This points to a relativistic scaling structure that may persist beyond the critical fixed point, despite the inequivalence of the temporal and spatial sectors in the microscopic action.

Despite its irrelevance, $h$ has a pronounced effect on the crossover towards asymptotic criticality. Our estimate for its associated correction-to-scaling exponent is remarkably small but sizable, with $\omega_h=0.0330$ for the classical model in $d=3$ and $\omega_h=0.0353$ for the quantum model in $d=2$. Thus, $h$ flows only slowly towards zero, potentially producing substantial deviations from conventional $O(N)$ scaling over an extended range of length scales. The asymptotic regime may consequently be reached only very close to criticality, making these corrections important for interpreting numerical simulations and experiments.

Overall, our results establish the irrelevance of the additional kinetic coupling for both classical and quantum phase transitions, while showing that its weakly irrelevant nature can have important consequences for the observable critical behavior. These findings thus provide a more complete picture of the critical properties of a broad class of physical systems characterized by interlocked internal and spatial symmetries.

Several interesting questions remain open and provide directions for future work. First, the quantitative discrepancy between our results and those of Ref.~\cite{Adzhemyan2023} calls for further investigation. Resolving it will likely require higher-order derivative expansions or complementary approaches, together with more extensive numerical data. In particular, Monte Carlo studies could provide an important independent test, building on the first results of Ref.~\cite{Itou2025}.

A natural extension would be to investigate the effects of reducing the rotational symmetry to a discrete subgroup. In the classical $O(d=3)$ model, for instance, a cubic perturbation is relevant, whereas the additional kinetic coupling considered here is reported to be irrelevant~\cite{Aharony1973c}. It would be interesting to determine how their interplay affects the corrections to scaling. In the quantum $O(d=2)$ model, by contrast, the cubic coupling is irrelevant~\cite{Leonard2015}. Since irrelevant couplings can nevertheless modify critical exponents on the symmetry-broken side of the transition~\cite{Leonard2015}, it would be worthwhile to explore whether the kinetic coupling $h$ influences this behavior as well.

Another direction concerns the breaking of the $\mathds Z_2$ symmetry, which would allow additional terms such as $\rho\partial_i\phi_i$. Such terms may be relevant in a variety of settings, including ferroelectric systems~\cite{Hinshaw1988,Chandra2017}, where they have been predicted to drive the emergence of modulated phases, as well as models with Lorentz-symmetry breaking, such as the bosonic sector of the Thirring-Yukawa model~\cite{Janssen2012}. They may also arise in non-centrosymmetric crystals or in the presence of an external magnetic field, where inversion or time-reversal symmetry, respectively, is broken. Extending the FRG framework developed in the present work would provide a means to investigate the resulting critical behavior.

Finally, it would be interesting to explore the limit $h\to -1$, which favors longitudinal over transverse fluctuations. In contrast to the limit $h\to\infty$, which suppresses longitudinal fluctuations, this regime may lead to effective one-dimensional dynamics. The resulting fixed-point structure and critical bcehavior in the limit $h\to -1$ remain to be investigated.

\begin{acknowledgments}

We are grateful to Bertrand Delamotte, Bilal Hawashin, and Nicolás Wschebor for discussions. L.C. also acknowledges discussions with Rafael Fernandes and Elio K\"onig on a related project. M.F., K.F., and L.C. were funded by the European Union (ERC-2023-STG, Project No. 101115758 - QuantEmerge). Views and opinions expressed are, however, those of the author only and do not necessarily reflect those of the European Union or the European Research Council Executive Agency. Neither the European Union nor the granting authority can be held responsible for them.
L.C. was also supported by the Simons Foundation SFI-MPS-NFS-00006741-11, and the Deutsche Forschungsgemeinschaft (DFG, German Research Foundation) under Germany’s Excellence Strategy–EXC–2111–390814868 and TRR 360 – 492547816.
MMS was funded by Deutsche Forschungsgemeinschaft (DFG) within Project-ID 277146847, SFB 1238 (project C02), and the DFG Heisenberg programme (Project-ID 452976698).
\end{acknowledgments}

\bibliography{main.bbl}

\onecolumngrid
\appendix

\section{Threshold functions for the effective potential}
\label{app:thresholds}

In this Appendix we give the analytical expressions for the loop integrals for each of the regulators considered in the main text, both for arbitrary shape function and the linear shape function. These are expressed in terms of threshold functions.

\subsection{Symmetric regulator}

For the symmetric regulator, the threshold functions are defined as
\begin{align}
    I_G(\omega,\eta_\phi) &= \frac 12 k^{-d} \int_\mathbf{p} \mathbf p^2 (\partial_t r_k - \eta_\phi r_k) \frac{1}{A(\mathbf p^2,\omega)}, \\
    I_R(\omega_1,\omega_2,h,\eta_\phi) &= \frac 12 k^{-d} \int_\mathbf{p} \mathbf p^2 (\partial_t r_k - \eta_\phi r_k) \frac{A(\mathbf p^2,\omega_1) + B(\mathbf p^2,\omega_1+\omega_2,h)}{A(\mathbf p^2,\omega_1) B(\mathbf p^2,\omega_1+\omega_2,h) + C(\mathbf p^2,\omega_2,h) \sin^2 \theta}, \\
    I_h &= 0,
\end{align}
where we have introduced the coefficient functions
\begin{equation}
    A(\mathbf p^2,\omega) = \mathbf p^2 (1 + r_k) + \omega, \quad B(\mathbf p^2,\omega,h) = \mathbf p^2 (1 + h + r_k) + \omega, \quad C(\mathbf p^2,\omega,h) = h \mathbf p^2 \omega.
\end{equation}
For the linear shape function, we find
\begin{align}
    I_G(\omega,\eta_\phi) &= \frac{4 v_d}{d}
    \left(
    1 - \frac{\eta_\phi}{d + 2}
    \right), \\
    I_R(\omega_1,\omega_2,h,\eta_\phi) &= -\frac{v_{d-1}}{2\pi} \int_0^1 \diff y y^{d/2} \int_0^\pi \diff \theta \sin^{d-2} \theta
    \left[
    \left(
    \frac 1y - 1
    \right)
    \eta_\phi - \frac 2y
    \right]
    \frac{yh + 2(1 + \omega_1) + \omega_2}{(1 + \omega_1)(1 + yh + \omega_1 + \omega_2) + y h \omega_2 \sin^2 \theta},
\end{align}
where we have introduced the volume factor $v_d^{-1} = 2^{d+1} \pi^{d/2} \Gamma(\tfrac d2)$.

\subsection{Phase-space adjusted regulator}

For the phase-space-adjusted regulator, we obtain
\begin{align}
    I_G(\omega,\eta_\phi) &= \frac 12 k^{-d} \int_\mathbf{p} \mathbf p^2 (\partial_t r_k - \eta_\phi r_k) \frac{1}{A(\mathbf p^2,\omega)}, \\
    I_R(\omega_1,\omega_2,h,\eta_\phi) &= \frac 12 k^{-d} \int_\mathbf{p} \mathbf p^2 (\partial_t r_k - \eta_\phi r_k) \frac{A(\mathbf p^2,\omega_1) + \tilde B(\mathbf p^2,\omega_1+\omega_2,h)}{A(\mathbf p^2,\omega_1) \tilde B(\mathbf p^2,\omega_1+\omega_2,h) + \tilde C(\mathbf p^2,\omega_2,h) \sin^2 \theta}, \\
    I_h(\omega_1,\omega_2,h,\eta_\phi,\eta_h) &= \frac 12 h k^{-d} \int_\mathbf{p} \mathbf p^2 (\partial_t r_k - \eta_h r_k) \frac{A(\mathbf p^2,\omega_1) + \omega_2 \sin^2 \theta}{A(\mathbf p^2,\omega_1) \tilde B(\mathbf p^2,\omega_1+\omega_2,h) + \tilde C(\mathbf p^2,\omega_2,h) \sin^2 \theta}
\end{align}
with modified coefficient functions
\begin{equation}
    \tilde B(\mathbf p^2,\omega,h) = \mathbf p^2(1 + h)(1 + r_k) + \omega, \quad \tilde C(\mathbf p^2,\omega,h) = h \mathbf p^2 (1 + r_k) \omega,
\end{equation}
and where $2 \rho \cos^2 \theta = \chi_p$ as before. For the linear shape function, the threshold functions evaluate to
\begin{align}
    I_G(\omega,\eta_\phi) &= \frac{4 v_d}{d}
    \left(
    1 - \frac{\eta_\phi}{d + 2}
    \right), \\
    I_R(\omega_1,\omega_2,h,\eta_\phi) &= \frac{2 v_{d-1}}{\pi d}
    \left(
    1 - \frac{\eta_\phi}{d + 2}
    \right)
    \int_0^\pi \diff \theta \sin^{d-2} \theta \frac{2(1 + \omega_1) + h + \omega_2}{(1 + \omega_1)(1 + h + \omega_1 + \omega_2) + h \omega_2 \sin^2 \theta}, \\
    I_h(\omega_1,\omega_2,h,\eta_\phi,\eta_h) &= \frac{2 v_{d-1}}{\pi d} h
    \left(
    1 - \frac{\eta_h}{d + 2}
    \right)
    \int_0^\pi \diff \theta \sin^{d-2} \theta \frac{1 + \omega_1 + \omega_2 \sin^2 \theta}{(1 + \omega_1)(1 + h + \omega_1 + \omega_2) + h \omega_2 \sin^2 \theta}.
\end{align}

\subsection{Covariant regulator}

For the covariant regulator, the threshold functions are defined as
\begin{align}
    I_G(\omega,\eta_\phi,z) &= \frac 12 k^{-d-1} \int_p 
    \left[
    p^2 (\partial_t r_k - \eta_\phi r_k) + 2(z-1) p_0^2 r_k
    \right]
    \frac{1}{A(p^2,\omega)}, \\
    I_R(\omega_1,\omega_2,h,\eta_\phi,z) &= \frac 12 k^{-d-1} \int_p 
    \left[
    p^2 (\partial_t r_k - \eta_\phi r_k) + 2(z-1) p_0^2 r_k
    \right] \times \nonumber\\
    &\phantom{\frac 12 k^{-d-1} \int_p}\times
    \frac{A(p^2,\omega_1) + B(p^2,\omega_1+\omega_2,h \sin^2 \varphi)}{A(p^2,\omega_1) B(p^2,\omega_1+\omega_2,h \sin^2 \varphi) + C(p^2,\omega_2,h \sin^2 \varphi) \sin^2 \theta}, \\
    I_h &= 0,
\end{align}
where we have abbreviated $p^2 = p_0^2 + \mathbf p^2$ and introduced the polar angle between frequencies and spatial momenta
\begin{equation}
    p_0^2 = p^2 \cos^2 \varphi, \quad \mathbf p^2 = p^2 \sin^2 \varphi.
\end{equation}
The coefficient functions $A$, $B$ and $C$ take the same functional form as for the symmetric regulator. For the linear shape function, we find
\begin{align}
    I_G(\omega,\eta_\phi) &= \frac{4 v_{d+1}}{d+1} \frac{1}{1+\omega}
    \left[
    \frac{1 + 2z + d(d+4)}{(d+1)(d+3)} - \frac{\eta_\phi}{d+3}
    \right], \\
    I_R(\omega_1,\omega_2,h,\eta_\phi) &= -\frac{v_{d-1}}{(2\pi)^2} \int_0^\infty \diff y y^\frac{d+1}{2} \int_0^\pi \diff \varphi \sin^{d-1} \varphi \int_0^\pi \diff \theta \sin^{d-2} \theta
    \left[
    \left(
    \frac 1y - 1
    \right)
    [\eta_\phi - 2(z-1) \cos^2 \varphi] - \frac 2y 
    \right] \times \nonumber\\
    &\phantom{-\frac{v_{d-1}}{(2\pi)^2} \int_0^\infty \diff y y^\frac{d+1}{2}}
    \times \frac{2(1+\omega_1) + \omega_2 + y h \sin^2 \varphi}{(1 + \omega_1)(1 + y h \sin^2 \varphi + \omega_1 + \omega_2) + y h \omega_2 \sin^2 \varphi \sin^2 \theta}.
\end{align}

\subsection{Spectrally adjusted regulator}

For the spectrally adjusted regulator, the threshold functions are defined as
\begin{align}
    I_G(\omega,\eta_\phi,z) &= \frac 12 k^{-d-1} \int_p 
    \left[
    p^2 (\partial_t r_k - \eta_\phi r_k) + 2(z-1) p_0^2 r_k
    \right]
    \frac{1}{A(p^2,\omega)}, \\
    I_R(\omega_1,\omega_2,h,\eta_\phi,z) &= \frac 12 k^{-d-1} \int_p 
    \left[
    p^2 (\partial_t r_k - \eta_\phi r_k) + 2(z-1) p_0^2 r_k
    \right] \times \nonumber\\
    &\phantom{\frac 12 k^{-d-1} \int_p}
    \times \frac{A(p^2,\omega_1) + \tilde B(p^2,\omega_1+\omega_2,h \sin^2 \varphi)}{A(p^2,\omega_1) \tilde B(p^2,\omega_1+\omega_2,h \sin^2 \varphi) + \tilde C(p^2,\omega_2,h \sin^2 \varphi) \sin^2 \theta}, \\
    I_h(\omega_1,\omega_2,h,\eta_\phi,\eta_h,z) &= \frac 12 h k^{-d-1} \int_p p^2 \sin^2 \varphi (\partial_t r_k - \eta_h r_k) \times \nonumber\\
    &\phantom{\frac 12 h k^{-d-1} \int_p}
    \times \frac{A(p^2,\omega_1) + \omega_2 \sin^2 \theta}{A(p^2,\omega_1) \tilde B(p^2,\omega_1+\omega_2,h \sin^2 \varphi) + \tilde C(p^2,\omega_2,h \sin^2 \varphi) \sin^2 \theta},
\end{align}
again with the previously defined coefficient functions. For the linear shape function, the threshold functions evaluate to
\begin{align}
    I_G(\omega,\eta_\phi,z) &= \frac{4 v_{d+1}}{d+1} \frac{1}{1+\omega}
    \left[
    \frac{1 + 2z + d(d+4)}{(d+1)(d+3)} - \frac{\eta_\phi}{d+3}
    \right], \\
    I_R(\omega_1,\omega_2,h,\eta_\phi,z) &= \frac{v_{d-1}}{\pi d} \frac{1}{d+1} \int_0^\pi \diff \varphi \sin^{d-1} \varphi \int_0^\pi \diff \theta \sin^{d-2} \theta
    \left(
    1 - \frac{\eta_\phi}{d + 3} + 2 \cos^2 \varphi \frac{z-1}{d+3}
    \right) \times \nonumber\\
    &\phantom{\frac{v_{d-1}}{\pi d} \frac{1}{d+1} \int}
    \times \frac{2(1 + \omega_1) + h \sin^2 \varphi + \omega_2}{(1 + \omega_1)(1 + h \sin^2 \varphi + \omega_1 + \omega_2) + h \omega_2 \sin^2 \varphi \sin^2 \theta}, \\
    I_h(\omega_1,\omega_2,h,\eta_\phi,\eta_h,z) &= \frac{v_{d-1}}{\pi^2} \frac{1}{d+1} h
    \left(
    1 - \frac{\eta_h}{d + 3}
    \right)
    \int_0^\pi \diff \varphi \sin^{d-1} \varphi \int_0^\pi \diff \theta \sin^{d-2} \theta \times \nonumber\\
    &\phantom{\frac{v_{d-1}}{\pi^2} \frac{1}{d+1} \int}
    \times \frac{1 + \omega_1 + \omega_2 \sin^2 \theta}{(1 + \omega_1)(1 + h \sin^2 \varphi + \omega_1 + \omega_2) + h \omega_2 \sin^2 \varphi \sin^2 \theta}.
\end{align}

\section{Projections for the anomalous dimensions}
\label{app:proj-andim}

\subsection{Na\"ive projection}

Here, we show how the projector for $Z_\phi$ is usually derived for $O(N)$ models and why na\"ive interpolation to $O(d)$ models fails. Following the argument of Refs.~\cite{Wetterich1993a,Tetradis1994}, the two-point function of the $O(N)$ model reads, up to second order in momentum
\begin{equation}
    \Gamma_{k,ij}^{(2)}(p) = 
    \left[
    U_k^\prime(\bar\rho) + Z_\phi(\bar\rho) p^2
    \right]
    \delta_{ij} + 
    \left[
    U_k^{\prime\prime}(\bar\rho) + \frac 12 Y_\phi(\bar\rho) p^2
    \right]
    \phi_i \phi_j + \mathcal O(p^4).
\end{equation}
Na\"ively, we set $Y_\phi(\bar\rho) = 0$, since we are working in the LPA' truncation, and obtain $Z_\phi(\bar\rho)$ from the projection of the two-point function at constant field
\begin{equation}
    Z_\phi(\bar\rho) = \frac 1N \frac{\partial}{\partial p^2} \delta_{ij} \left(
    \Gamma_{k,0}^{(2)} 
    \right)_{ij}
    \Big\vert_{p=0}.
\end{equation}
However, while we can set $Y_\phi(\bar\rho) = 0$ in the UV, the right hand side of the Wetterich equation 
can include contributions that spoil this projection, because finite contributions to the terms omitted in LPA' will be generated during the flow. Applying the projection to the full two-point function, including $Y_\phi(\bar\rho)$, yields
\begin{equation}
    \frac 1N \frac{\partial}{\partial p^2} \delta_{ij}
    \left(
    \Gamma_{k,0}^{(2)} 
    \right)_{ij}
    \Big\vert_{p=0} = Z_\phi(\bar\rho) + \frac 1N \bar\rho Y_\phi(\bar\rho),
\end{equation}
which means that our truncation does not project cleanly onto $Z_\phi$, but is contaminated by flow contributions beyond LPA'. To remedy this issue, we can formally expand the two-point function in a complete basis of rank-two symmetric tensors. In the $O(N)$ model, this basis is two-dimensional and consists of the longitudinal and transverse projectors with respect to a constant field configuration
\begin{equation}
    P_{\phi,ij}^\mathrm{L} = \frac{\phi_i \phi_j}{2 \bar\rho}, \quad P_{\phi,ij}^\mathrm{T} = 1 - P_{\phi,ij}^\mathrm{L}.
\end{equation}
Projecting onto the transverse components of the field then gives
\begin{equation}
    \frac{\partial}{\partial p^2} P_{\phi,ij}^\mathrm{T} 
    \left(
    \Gamma_{k,0}^{(2)} 
    \right)_{ij}
    \Big\vert_{p=0} = (N-1)Z_\phi(\bar\rho),
\end{equation}
which cleanly projects onto $Z_\phi$. However, in the $O(d)$ model, this projection already fails at the LPA' level
\begin{equation}\label{eq:transvProj-Od}
    \frac{\partial}{\partial p^2} P_{\phi,ij}^\mathrm{T}
    \left(
    \Gamma_{k,0}^{(2)} 
    \right)_{ij}
    \Big\vert_{p=0} = (d-1)Z_\phi(\bar\rho) + \frac 1d Z_h(\bar\rho),
\end{equation}
with additional contributions arising from higher-order derivative expansion terms. The reason lies in an enlarged tensor basis, which now includes symmetric rank-two tensors built from momenta, as well as combinations of field and momenta. To construct a consistent projection scheme, we need to take the full space of symmetric rank-two tensors into account. The derivation of a consistent projection scheme for the $O(d)$ anomalous dimensions is presented in the remainder of this Appendix.

\subsection{Classical model}

Since the two-point function evaluated at constant field $\Gamma_{k,0}^{(2)}(p)$ is a symmetric rank-two tensor under $O(d) \times \mathds Z_2$, it naturally decomposes into
\begin{equation}
    \left(
    \Gamma_{k,0}^{(2)} 
    \right)_{ij} 
    (p) = \sum_\alpha \Gamma_\alpha^{(2)}(\bar\rho,p^2,\chi_p)\, T_{ij}^\alpha(p,\phi),
    \quad \chi_p = (p_i \phi_i)^2, 
\end{equation}
where $\Gamma_\alpha^{(2)}$ are scalar coefficients that only depend on invariants of $O(d) \times \mathds Z_2$ and
\begin{equation}
    T_{ij}^\alpha \in 
    \left\{
    \delta_{ij}, p_i p_j, \phi_i \phi_j, (p_i \phi_j + \phi_i p_j) p_k \phi_k
    \right\},
\end{equation}
denotes a basis of symmetric rank-two tensors built from the $O(d)$ vectors $p$ and $\phi$. Treating the momentum-dependent invariants $p^2$ and $\chi_p$ as independent variables, the coefficient functions can be expanded in a power series
\begin{equation}
    \Gamma_\alpha^{(2)}(\bar\rho,p^2,\chi_p) = \sum_{m,n} Z_{\alpha,mn}(\bar\rho) (p^2)^m \chi_p^n, \quad Z_{\alpha,mn}(\bar\rho) = \frac{1}{m! n!} \frac{\partial^m}{\partial (p^2)^m} \frac{\partial^n}{\partial \chi_p^n} \Gamma_\alpha^{(2)}(\bar\rho,p^2,\chi_p) \bigg\vert_{p^2 = \chi_p = 0}.
\end{equation}
With that, the full two-point function reads
\begin{equation}
    \left(
    \Gamma_{k,0}^{(2)} 
    \right)_{ij}
    (p) = \sum_{\alpha,m,n} Z_{\alpha,mn}(\bar\rho) (p^2)^m \chi_p^n\, T_{ij}^\alpha(p,\phi),
\end{equation}
We only consider terms of order $\mathcal O(p^2)$, capturing all momentum-dependent operators to second order in the derivative expansion
\begin{align}
    \mathcal D_{ab,ij} \equiv \frac{\partial^2}{\partial p_a \partial p_b} \left(
    \Gamma_{k,0}^{(2)} 
    \right)_{ij}
    \Big\vert_{p=0} &= 2
    \left[
    Z_{1,10}(\bar\rho) \delta_{ab} + Z_{1,01}(\bar\rho) \phi_a \phi_b
    \right]
    \delta_{ij} + Z_{2,00}(\bar\rho) (\delta_{ai}\delta_{bj} + \delta_{aj}\delta_{bi}) \nonumber\\
    &+ 2 
    \left[
    Z_{3,10}(\bar\rho) \delta_{ab} + Z_{3,01}(\bar\rho) \phi_a \phi_b
    \right]
    \phi_i \phi_j + Z_{4,00}(\bar\rho) [\phi_a (\phi_i \delta_{bj} + \phi_j \delta_{bi}) + \phi_b (\phi_i \delta_{aj} + \phi_j \delta_{ai})].
\end{align}
Within the LPA' truncation, the two-point function reads
\begin{equation}
    \left(
    \Gamma_{k,0}^{(2)} 
    \right)_{ij}
    = 
    \left[
    U_k^\prime(\bar\rho) + Z_\phi(\bar\rho) p^2
    \right]
    \delta_{ij} + Z_h(\bar\rho) p_i p_j + U_k^{\prime\prime}(\bar\rho) \phi_i \phi_j.
\end{equation}
Hence, it is sufficient to derive projectors onto the renormalization factors $Z_\phi(\bar\rho) \equiv Z_{1,10}(\bar\rho)$ and $Z_h(\bar\rho) \equiv Z_{2,00}(\bar\rho)$, which only appear in transverse tensor structures in $\mathcal D_{ab,ij}$. To remove longitudinal contributions, we use the transverse projector
\begin{align}
    P_{ab}^T P_{ij}^T \mathcal D_{ab,ij} &= 2 (d - 1)^2 Z_\phi(\bar\rho) + 2 (d - 1) Z_h(\bar\rho), \\
    P_{ai}^T P_{bj}^T \mathcal D_{ab,ij} &= 2 (d - 1) Z_\phi(\bar\rho) + d (d - 1) Z_h(\bar\rho),
\end{align}
from which we can read off the projectors
\begin{align}
    Z_\phi(\bar\rho) &= \frac{1}{(d^2 - d - 2)(d - 1)}
    \left(
    \frac d2 P_{ab}^T P_{ij}^T - P_{ai}^T P_{bj}^T
    \right)
    \mathcal D_{ab,ij} \equiv \Pi_{ab,ij}^\phi \mathcal D_{ab,ij}, \label{eq:proj-phi}\\
    Z_h(\bar\rho) &= \frac{1}{(d^2 - d - 2)(d - 1)}
    \left[
    - P_{ab}^T P_{ij}^T + (d - 1) P_{ai}^T P_{bj}^T
    \right]
    \mathcal D_{ab,ij} \equiv \Pi_{ab,ij}^h \mathcal D_{ab,ij}. \label{eq:proj-h}
\end{align}
Projectors for $\mathcal O(p^2)$ operators that have been neglected in our LPA' truncation can be constructed by considering the longitudinal subspace of $\mathcal D_{ab,ij}$. Operators beyond second order in the derivative expansion arise from projection of tensors with more than two indices associated with momentum derivatives. 

\subsection{Quantum model}

For the quantum model, the additional temporal derivative in the effective action induces a frequency dependence of the two-point function. However, since frequencies transform trivially under spatial rotations, no additional tensor structures appear in the expansion of the two-point function. More precisely, the frequency-dependence only enters the coefficient functions via the invariant $\omega^2$, giving
\begin{equation}
    \left(
    \Gamma_{k,0}^{(2)} 
    \right)_{ij} (p)
    = \sum_\alpha \Gamma_\alpha^{(2)}(\bar\rho,\omega^2, \mathbf p^2,\chi_p)\, T_{ij}^\alpha(\mathbf p,\phi),
    \quad \chi_p = (p_i \phi_i)^2, 
\end{equation}
with the same tensor basis $T^\alpha$ as before. For the LPA' truncation we have
\begin{equation}
    \left(
    \Gamma_{k,0}^{(2)} 
    \right)_{ij} (p)
    = 
    \left[
    U_k^\prime(\bar\rho) + Z_{\phi,\parallel}(\bar\rho) \omega^2 + Z_{\phi,\perp}(\bar\rho) p^2
    \right]
    \delta_{ij} + Z_h(\bar\rho) p_i p_j + U_k^{\prime\prime}(\bar\rho) \phi_i \phi_j.
\end{equation}
Due to the additional frequency-dependence, the full expansion then reads
\begin{equation}
    \left(
    \Gamma_{k,0}^{(2)} 
    \right)_{ij} (p)
    = \sum_{\alpha,m,n,o} Z_{\alpha,mno}(\bar\rho) (\omega^2)^m (p^2)^n \chi_p^o\, T_{ij}^\alpha(p,\phi),
\end{equation}
Hence, we find $\mathcal D_{ab,ij}$ as before and
\begin{equation}
    \mathcal E_{ij} \equiv \frac{\partial^2}{\partial \omega^2} \left(
    \Gamma_{k,0}^{(2)} 
    \right)_{ij}
    \Big\vert_{\omega = \mathbf p = 0} = 2 Z_{1,100}(\bar\rho) \delta_{ij} + 2 Z_{3,100}(\bar\rho) \phi_i \phi_j.
\end{equation}
We want to extract $Z_{\phi,\parallel}(\bar\rho) \equiv Z_{1,100}(\bar\rho)$, which we obtain from
\begin{equation}
    P_{ij}^T \mathcal E_{ij} = 2 (d - 1) Z_{\phi,\parallel} \to Z_{\phi,\parallel} = \frac 12 \frac{1}{d - 1} P_{ij}^T \mathcal E_{ij}.
\end{equation}
Because terms $\mathcal O(\omega \mathbf p, \omega \phi)$ are forbidden by symmetry, the rank-4 tensor describing frequency and momentum derivatives of the two-point function is block-diagonal in the derivative indices
\begin{equation}
    \begin{pmatrix}
        \mathcal E_{ij} & 0 \\
        0 & \mathcal D_{ab,ij}
    \end{pmatrix},
\end{equation}
such that the projectors onto the couplings $Z_{\phi,\perp} \equiv Z_{1,010}$ and $Z_h \equiv Z_{2,000}$ are the same as in the classical case.

\section{Flow equations for the anomalous dimensions}
\label{app:flow-andim}

\subsection{Tensor decomposition of the propagator}

Here, we specify the tensor decomposition of the propagator introduced in the FRG equations for the anomalous dimensions below Eq.~\eqref{eq:prop-decomp}. The coefficient functions $G_\alpha$ are obtained by inverting the regularized two-point function using the identity
\begin{equation}
    \left[
    \Gamma_{k}^{(2)}(q) + R_k(q)
    \right]
    _{ab}
    G_{k,bc}(q) = \delta_{ac}.
\end{equation}
Here, $q$ denotes spatial momentum in the classical case and frequency-momentum $(\nu,\mathbf q)$ in the quantum case. In agreement with the derivation in Ref.~\cite{Kalagov2026}, we obtain
\begin{align}
    G_1 &= \frac{1}{\Gamma_1}, \quad
    G_2 = \frac{-\Gamma_2 (\Gamma_1 + 2 \rho \Gamma_3) + 2 \rho \chi_q (\Gamma_4)^2}{g(\rho,\mathbf q^2,\chi_q)}, \\
    G_3 &= \frac{-\Gamma_3 (\Gamma_1 + \mathbf q^2 \Gamma_2) + \mathbf q^2 \chi_q (\Gamma_4)^2}{g(\rho,\mathbf q^2,\chi_q)}, \quad
    G_4 = \frac{-\Gamma_4 (\Gamma_1 + \chi_q \Gamma_4) + \Gamma_2 \Gamma_3}{g(\rho,\mathbf q^2,\chi_q)},
\end{align}
with
\begin{equation}
    g(\rho,\mathbf q^2,\chi_q) = \Gamma_1 [(\Gamma_1)^2 + \Gamma_1 (\mathbf q^2 \Gamma_2 + 2 \rho \Gamma_3 + 2 \chi_q \Gamma_4) + (2 \mathbf q^2 \rho - \chi_q) (\Gamma_2 \Gamma_3 - \chi_q^2 \Gamma_4)],
\end{equation}
where $\Gamma_\alpha$ denote the projections of the regularized two-point function onto the respective tensor basis elements. We quote the $\Gamma_\alpha$ for the regulators used in the main text. For all regulators, we have
\begin{equation}
    \Gamma_3 = 2 \bar\rho U_k''(\bar\rho), \quad \Gamma_4 = 0.
\end{equation}
For the symmetric regulator \eqref{eq:sym-reg} we additionally find
\begin{equation}
    \Gamma_1 = Z_\phi \mathbf p^2 [1 + r_k(\mathbf p^2)] + U_k'(\bar\rho), \quad \Gamma_2 = Z_h.
\end{equation}
For the phase-space-adjusted regulator \eqref{eq:adj-reg}, we find
\begin{equation}
    \Gamma_1 = Z_\phi \mathbf p^2 [1 + r_k(\mathbf p^2)] + U_k'(\bar\rho), \quad \Gamma_2 = Z_h [1 + r_k(\mathbf p^2)].
\end{equation}
For the covariant regulator \eqref{eq:cov-regQ} we find
\begin{equation}
    \Gamma_1 = (Z_{\phi,\parallel} \omega^2 + Z_{\phi,\perp} \mathbf p^2) [1 + r_k(\omega^2/v_\phi^2 + \mathbf p^2)] + U_k'(\bar\rho), \quad \Gamma_2 = Z_h.
\end{equation}
For the quantum, phase-space-adjusted regulator \eqref{eq:adj-regQ}, we find
\begin{equation}
    \Gamma_1 = (Z_{\phi,\parallel} \omega^2 + Z_{\phi,\perp} \mathbf p^2) [1 + r_k(\omega^2/v_\phi^2 + \mathbf p^2)] + U_k'(\bar\rho), \quad \Gamma_2 = Z_h [1 + r_k(\omega^2/v_\phi^2 + \mathbf p^2)].
\end{equation}

\subsection{Classical model}

To obtain the flow equations for the renormalization factors, we apply the projectors \eqref{eq:proj-phi} and \eqref{eq:proj-h} to the flow of the two-point function at constant field, which produce the flow equations
\begin{equation}
    \partial_t Z_\phi(\bar\rho) = \Pi_{ab,ij}^\phi \frac{\partial^2 }{\partial p_a \partial p_b} \partial_t 
    \left(
    \Gamma_{k,0}^{(2)} 
    \right)_{ij}
    \Big\vert_{p=0}, \quad \partial_t Z_h(\bar\rho) = \Pi_{ab,ij}^h \frac{\partial^2 }{\partial p_a \partial p_b} \partial_t
    \left(
    \Gamma_{k,0}^{(2)} 
    \right)_{ij}
    \Big\vert_{p=0}.
\end{equation}
The flow of the two-point function is derived by taking two functional derivatives of the Wetterich equation. We find
\begin{equation}\label{eq:flow-twopoint}
    \partial_t \Gamma_{k,0}^{(2)}(p) = \int_q \partial_t R_k(q) \Tr
    \left[
    G_k(q) U_k^{(3)} G_k(q+p) U_k^{(3)} G_k(q) - \frac 12 G_k(q) U_k^{(4)} G_k(q)
    \right],
\end{equation}
where we have introduced the propagator at constant field $\phi$
\begin{equation}
    G_k(p) = \left[
    \Gamma_k^{(2)}(p) + R_k(p)
    \right]^{-1} \Big\vert_{\phi = \mathrm{const.}},
\end{equation}
and the local three-point function, also at constant field
\begin{equation}
    U_{k,abc}^{(3)} = \left(\delta_{ab}\phi_c + \delta_{ac}\phi_b + \delta_{bc}\phi_a\right) \,U_k^{\prime\prime}(\bar\rho) + \phi_a \phi_b \phi_c U_k^{\prime\prime\prime}(\bar\rho) \Big\vert_{\phi = \mathrm{const.}}. 
\end{equation}
Since the second term in Eq. \eqref{eq:flow-twopoint}, which contains the local four-point function at constant field $U_k^{(4)}$, does not depend on external momentum, it does not contribute to the flow of the momentum-dependent part of the two-point function and can be safely ignored hereafter. The first term can be rewritten in terms of the modified scale derivative
\begin{equation}
    \tilde\partial_t = \Tr \int_p \partial_t R_k(p) \frac{\delta}{\delta R_k(p)},
\end{equation}
and subsequently expanded to quadratic order in external momentum
\begin{align}
    \frac{\partial^2 }{\partial p_a \partial p_b} \partial_t\Gamma_{k,0}^{(2)} \Big\vert_{p=0}
    &= -\frac 14 \frac{\partial^2 }{\partial p_a \partial p_b} \int_q \tilde\partial_t \Tr
    \left[
    G_k(q) U_k^{(3)} G_k(q+p) U_k^{(3)}
    \right] \Big\vert_{p = 0} \\
    &= \frac 14 \frac{\partial^2 }{\partial p_a \partial p_b} \int_q \tilde\partial_t \Tr
    \left[
    \frac{\partial}{\partial q_i}G_k(q) U_k^{(3)} \frac{\partial}{\partial q_j} G_k(q) U_k^{(3)}
    \right]
    p_i p_j \\
    &= \frac 12 \int_q \tilde\partial_t \Tr
    \left[
    \frac{\partial}{\partial q_a}G_k(q) U_k^{(3)} \frac{\partial}{\partial q_b} G_k(q) U_k^{(3)}
    \right]
\end{align}
We make further progress by decomposing the propagator into the tensor basis
\begin{equation}
    G_k(q,\phi) = G_\alpha(\bar\rho,q^2,\chi_q) T^\alpha(q,\phi),
\end{equation}
giving
\begin{equation}\label{eq:prop-decomp}
    \frac{\partial^2 }{\partial p_a \partial p_b} \partial_t
    \left(
    \Gamma_{k,0}^{(2)} 
    \right)_{ij}
    \Big\vert_{p=0}
    = \frac 12 \int_q \tilde\partial_t 
    \left\{
    \frac{\partial G_\alpha}{\partial q_a} \frac{\partial G_\beta}{\partial q_b} T_{cd}^\alpha U_{k,dei}^{(3)} T_{ef}^\beta U_{k,fcj}^{(3)} + G_\alpha G_\beta \frac{\partial T_{cd}^\alpha}{\partial q_a} U_{k,dei}^{(3)} \frac{\partial T_{ef}^\beta}{\partial q_b} U_{k,fcj}^{(3)}
    \right\}.
\end{equation}
From there, we apply the projectors, giving
\begin{align}
    \partial_t Z_\phi(\bar\rho) &= \frac{4 \bar\rho\, [U_k^{\prime\prime}(\bar\rho)]^2}{d^2 - 1} \tilde\partial_t \int_q 
    \mathbf q^2 \sin^2 \theta \,\times \nonumber\\
    &\phantom{= d^2 - 1}
    \times \Bigg\{
    \frac{\partial G_1}{\partial \mathbf q^2}
    \left[
     (d + 1)
    \left(
    \frac{\partial G_1}{\partial \mathbf q^2} + \mathbf q^2 \cos^2 \theta \frac{\partial G_2}{\partial \mathbf q^2} + 2 \bar\rho \frac{\partial G_3}{\partial \mathbf q^2} + 4 \bar\rho \mathbf q^2 \cos^2 \theta \frac{\partial G_4}{\partial \mathbf q^2}
    \right)
    + \mathbf q^2 \sin^2 \theta \frac{\partial G_2}{\partial \mathbf q^2}
    \right] \nonumber\\
    &\phantom{= d^2 - 1}
    + 2 \mathbf q^2 \sin^2 \theta 
    \left[
    \mathbf q^2 \cos^2 \theta
    \left(
    \frac{\partial G_2}{\partial \mathbf q^2}
    \right)^2 
    + 2 \mathbf q^2 \bar\rho^2 \cos^2 \theta
    \left(
    \frac{\partial G_4}{\partial \mathbf q^2}
    \right)^2
    + \bar\rho \frac{\partial G_2}{\partial \mathbf q^2}
    \left(
    \frac{\partial G_3}{\partial \mathbf q^2} + 4 \mathbf q^2 \cos^2 \theta \frac{\partial G_4}{\partial \mathbf q^2}
    \right)
    \right]
    \Bigg\}, \label{eq:flow-Zphi}\\
    \partial_t Z_h(\bar\rho) &= \bar\rho [U_k^{\prime\prime}(\bar\rho)]^2 \tilde\partial_t \int_q \mathbf q^2
    \Bigg\{
    \frac{8 \mathbf q^2 \sin^4 \theta}{d^2 - 1}
    \bigg[
    \frac{\partial G_2}{\partial \mathbf q^2}
    \left(
    \frac{\partial G_1}{\partial \mathbf q^2}
    +2 \mathbf q^2 \cos^2 \theta \frac{\partial G_2}{\partial \mathbf q^2} + 2 \bar\rho \frac{\partial G_3}{\partial \mathbf q^2}
    \right) \nonumber\\
    &\phantom{\bar\rho [U_k^{\prime\prime}(\bar\rho)]^2 \tilde\partial_t \int_q \mathbf q^2}
    + 4 \mathbf q^2 \bar\rho \cos^2 \theta \frac{\partial G_4}{\partial \mathbf q^2}
    \left(
    2 \frac{\partial G_2}{\partial \mathbf q^2} + \bar\rho \frac{\partial G_4}{\partial \mathbf q^2}
    \right)
    \bigg]
    + \cos^2 \theta\,  (G_2 + 2 \bar\rho G_4)^2
    \Bigg\}, \label{eq:flow-Zh}
\end{align}
where we have introduced the angular variable $2 \bar\rho \mathbf q^2 \cos^2 \theta = \chi_q$. Within our ansatz, we approximate the renormalization factors by their value at the scale-dependent minimum of the effective potential $\bar\kappa$. The anomalous dimensions are then given by appropriately normalizing the flow equations \eqref{eq:flow-Zphi} and \eqref{eq:flow-Zh} 
\begin{equation}
    \eta_\phi = -\frac{\partial_t Z_\phi(\bar\rho)}{Z_\phi(\bar\rho)} \Bigg\vert_{\bar\rho = \bar\kappa}, 
    \quad 
    \eta_h = -\frac{\partial_t Z_h(\bar\rho)}{Z_h(\bar\rho)} \Bigg\vert_{\bar\rho = \bar\kappa}.
\end{equation}

\subsection{Quantum model}

Because the projectors for $Z_{\phi,\perp}$ and $Z_h$ agree in classical and quantum case, the integrands entering their flow equations are the same, but with coefficients $G_\alpha$ adapted to the quantum case. Additionally, we integrate over loop frequencies $\nu \in (-\infty,\infty)$. This defines the anomalous dimensions $\eta_{\phi,\perp}$ and $\eta_h$. For the flow of $Z_{\phi,\parallel}$ we have
\begin{align}
    \frac{\partial^2}{\partial \omega^2} \partial_t
    \left(
    \Gamma_{k,0}^{(2)} 
    \right)_{ij}
    \Big\vert_{\omega = \mathbf p=0} &= \frac 12 \int_{\nu,\mathbf q} \tilde\partial_t \Tr
    \left[
    \frac{\partial}{\partial \nu}G_k(q) U_k^{(3)} \frac{\partial}{\partial \nu} G_k(q) U_k^{(3)}\right] \\
    &= 2 \int_{\nu,\mathbf q} \tilde\partial_t 
    \left(
    \nu^2 \frac{\partial G_\alpha}{\partial \nu^2} \frac{\partial G_\beta}{\partial \nu^2} T_{cd}^\alpha U_{k,dei}^{(3)} T_{ef}^\beta U_{k,fcj}^{(3)}
    \right)
\end{align}
After evaluating the trace, we obtain
\begin{align}
    \partial_t Z_{\phi,\parallel}(\bar\rho) &= \frac{4 \bar\rho\, [U_k^{\prime\prime}(\bar\rho)]^2}{d - 1} \tilde\partial_t \int_{\nu,\mathbf q} 
    \nu^2 \,\times \nonumber\\
    &\phantom{= 4 \bar\rho\,}
    \times \Bigg\{
    \frac{\partial G_1}{\partial \nu^2}
    \left[
    (d - 1)
    \left(
    \frac{\partial G_1}{\partial \nu^2} + \mathbf q^2 \cos^2 \theta \frac{\partial G_2}{\partial \nu^2} + 2 \bar\rho \frac{\partial G_3}{\partial \nu^2} + 4 \bar\rho \mathbf q^2 \cos^2 \theta \frac{\partial G_4}{\partial \nu^2}
    \right)
    + \mathbf q^2 \sin^2 \theta \frac{\partial G_2}{\partial \nu^2}
    \right] \nonumber\\
    &\phantom{= 4 \bar\rho\,}
    + 2 \mathbf q^2 \sin^2 \theta 
    \left[
    \mathbf q^2 \cos^2 \theta
    \left(
    \frac{\partial G_2}{\partial \nu^2}
    \right)^2 
    + 2 \mathbf q^2 \bar\rho^2 \cos^2 \theta
    \left(
    \frac{\partial G_4}{\partial \nu^2}
    \right)^2
    + \bar\rho \frac{\partial G_2}{\partial \nu^2}
    \left(
    \frac{\partial G_3}{\partial \nu^2} + 4 \mathbf q^2 \cos^2 \theta \frac{\partial G_4}{\partial \nu^2}
    \right)
    \right]
    \Bigg\},
\end{align}
from which we infer the anomalous dimension
\begin{equation}
    \eta_{\phi,\parallel} = -\frac{\partial_t Z_{\phi,\parallel}(\bar\rho)}{Z_{\phi,\parallel}(\bar\rho)} \Bigg\vert_{\bar\rho = \bar\kappa}.
\end{equation}

\end{document}